\documentclass[twocolumn]{aastex6}
\usepackage{amsmath}
\newcommand{\appropto}{\mathrel{\vcenter{\offinterlineskip\halign{\hfil$##$\cr\propto\cr\noalign{\kern2pt}\sim\cr\noalign{\kern-2p
t}}}}}

\newcommand\earthsym{$_{\oplus}$}
\newcommand\sunsym{$_{\odot}$}
\newcommand\kepler{\emph{Kepler}}

\newcommand\epicno{EPIC~220383386}

\newcommand\distance{46}

\newcommand{\mse}{\mbox{m\,s$^{-1}$}}

\newcommand\pradb{1.70$^{+0.18}_{-0.15}$}
\newcommand\periodb{0.959641$^{+0.000011}_{-0.000012}$}
\newcommand\epochb{2457394.37454$\pm$0.00043}
\newcommand\depthb{294}
\newcommand\rponrsb{0.01744$^{+0.00170}_{-0.00089}$}
\newcommand\aonrsb{4.082$^{+0.464}_{-0.986}$}
\newcommand\impparb{0.47$^{+0.31}_{-0.32}$}
\newcommand\incb{83.4$^{+4.6}_{-7.7}$}
\newcommand\tdurb{1.622$^{+0.060}_{-0.074}$}
\newcommand\semiab{0.01815$\pm$0.00023}
\newcommand\pmassberr{5.02$\pm$0.38}
\newcommand\pdensberr{5.60$^{+2.15}_{-1.43}$}
\newcommand\rvampb{3.58$^{+0.25}_{-0.26}$}

\newcommand\pradc{3.01$^{+0.42}_{-0.28}$}
\newcommand\periodc{29.8454$\pm$0.0012}
\newcommand\epochc{2457394.9788$\pm$0.0012}
\newcommand\depthc{946}
\newcommand\rponrsc{0.0313$^{+0.0045}_{-0.0018}$}
\newcommand\aonrsc{40.323$^{+5.549}_{-12.622}$}
\newcommand\impparc{0.50$^{+0.31}_{-0.33}$}
\newcommand\incc{89.3$^{+0.5}_{-0.96}$}
\newcommand\tdurc{5.15$^{+0.26}_{-0.19}$}
\newcommand\semiac{0.1795$\pm$0.0023}
\newcommand\pmasscerr{9.80$^{+1.30}_{-1.24}$}
\newcommand\pdenscerr{1.97$^{+0.94}_{-0.59}$}
\newcommand\rvampc{2.23$^{+0.29}_{-0.28}$}

\newcommand\periodd{8.509$\pm$0.045}
\newcommand\tcepochd{2457806.07$^{+0.52}_{-0.50}$}
\newcommand\pmassderr{6.90$\pm$0.71}

\newcommand\semiad{0.07757$\pm$0.00027}

\slugcomment{Accepted in the Astronomical Journal}
\shorttitle{Masses of HD 3167 planets}
\shortauthors{Christiansen et al.}
\begin{document}


%\title{HD~3167 b is a hot rock with an observable atmosphere. HD~3167 c is probably interesting too.}
%\title{HD~3167: A bright, nearby G dwarf hosting a hot Super-Earth and a warm mini-Neptune.}
%\title{HD~3167: A new benchmark system for exoplanet atmosphere studies.}
\title{Three's Company: An additional non-transiting super-Earth in the bright HD~3167 system, and masses for all three planets.}

%% Use \author, \affil, plus the \and command to format author and affiliation 
%% information.  If done correctly the peer review system will be able to
%% automatically put the author and affiliation information from the manuscript
%% and save the corresponding author the trouble of entering it by hand.
%%
%% The \affil should be used to document primary affiliations and the
%% \altaffil should be used for secondary affiliations, titles, or email.

%% Authors with the same affiliation can be grouped in a single
%\author{Jessie L. Christiansen$^{1,2}$}
\author{Jessie L. Christiansen\altaffilmark{1,2}}
\author{Andrew Vanderburg\altaffilmark{16}}
\author{Jennifer Burt\altaffilmark{11}}
\author{B. J. Fulton\altaffilmark{4,8}}

\author{Konstantin Batygin\altaffilmark{6}}
\author{Bj\"{o}rn Benneke\altaffilmark{6}}
\author{John M. Brewer\altaffilmark{17}}
\author{David Charbonneau\altaffilmark{16}}
\author{David R. Ciardi\altaffilmark{1}}
\author{Andrew Collier Cameron\altaffilmark{22}}
\author{Jeffrey L. Coughlin\altaffilmark{18,19}}
\author{Ian~J.~M.~Crossfield\altaffilmark{3,13}}
\author{Courtney Dressing\altaffilmark{6,13}}
\author{Thomas P. Greene\altaffilmark{18}}
\author{Andrew W. Howard\altaffilmark{8}}
\author{David W. Latham\altaffilmark{16}}
\author{Emilio Molinari\altaffilmark{23,24}}
\author{Annelies Mortier\altaffilmark{22}}
\author{Fergal Mullally\altaffilmark{19}}
\author{Francesco Pepe\altaffilmark{21}}
\author{Ken Rice\altaffilmark{20}}
\author{Evan Sinukoff\altaffilmark{4,8}}
\author{Alessandro Sozzetti\altaffilmark{35}}
\author{Susan E. Thompson\altaffilmark{18,19}}
\author{St{\'e}phane Udry\altaffilmark{21}}
\author{Steven S. Vogt\altaffilmark{12}}

\author{Travis~S.~Barman\altaffilmark{5}}
\author{Natasha E. Batalha\altaffilmark{30}}
\author{Fran{\c c}ois Bouchy\altaffilmark{21}}
\author{Lars A. Buchhave\altaffilmark{29}}
\author{R. Paul Butler\altaffilmark{15}}
\author{Rosario Cosentino\altaffilmark{24}}
\author{Trent J. Dupuy\altaffilmark{7}}
\author{David Ehrenreich\altaffilmark{21}}
\author{Aldo Fiorenzano\altaffilmark{24}}
\author{Brad M. S. Hansen\altaffilmark{34}}
\author{Thomas Henning\altaffilmark{31}}
\author{Lea Hirsch\altaffilmark{9}}
\author{Bradford P. Holden\altaffilmark{12}}
\author{Howard T. Isaacson\altaffilmark{9}}
\author{John A. Johnson\altaffilmark{16}}
\author{Heather A. Knutson\altaffilmark{6}}
\author{Molly Kosiarek\altaffilmark{3}}
\author{Mercedes L{\'o}pez-Morales\altaffilmark{16}}
\author{Christophe Lovis\altaffilmark{21}}
\author{Luca Malavolta\altaffilmark{26,27}}
\author{Michel Mayor\altaffilmark{21}}
\author{Giuseppina Micela\altaffilmark{25}}
\author{Fatemeh Motalebi\altaffilmark{21}}
\author{Erik Petigura\altaffilmark{6}}
\author{David F. Phillips\altaffilmark{16}}
\author{Giampaolo Piotto\altaffilmark{26,27}}
%\author{Didier Queloz\altaffilmark{21}}
\author{Leslie A. Rogers\altaffilmark{33}}
\author{Dimitar Sasselov\altaffilmark{16}}
\author{Joshua E. Schlieder\altaffilmark{10}}
\author{Damien S{\'e}gransan\altaffilmark{21}}
\author{Christopher A. Watson\altaffilmark{28}}
\author{Lauren M. Weiss\altaffilmark{36,37}}
%\author{Michael W. Werner\altaffilmark{32}}
\email{jessie.christiansen@caltech.edu}
%\affil{$^1$NASA Exoplanet Science Institute, California Institute of Technology, M/S 100-22, 770 S. Wilson Ave, Pasadena, CA 91106, USA}
%\affil{$^2$jessie.christiansen@caltech.edu}
%\affil{$^3$Lunar \& Planetary Laboratory, University of Arizona, 1629 E. University Blvd., Tucson, AZ, USA}
%\affil{$^8$NASA Sagan Fellow}
%\affil{$^5$Institute for Astronomy, University of HawaiÔi at M\={a}noa, Honolulu, HI, USA}
%\affil{$^4$Geological and Planetary Sciences, California Institute of Technology, Pasadena, CA, USA}
%\affil{$^6$Astronomy Department, University of California, Berkeley, CA, USA}
%\affil{$^9$NSF Graduate Student Fellow}
%\affil{$^7$NSERC Postgraduate Research Fellow}

\altaffiltext{1}{NASA Exoplanet Science Institute, California Institute of Technology, M/S 100-22, 770 S. Wilson Ave, Pasadena, CA, USA}
\altaffiltext{2}{jessie.christiansen@caltech.edu}
\altaffiltext{3}{Department of Astronomy \& Astrophysics, University of California, Santa Cruz, CA, USA}
%\altaffiltext{8}{NASA Sagan Fellow}
\altaffiltext{4}{Institute for Astronomy, University of Hawai'i at M\={a}noa, Honolulu, HI, USA}
\altaffiltext{5}{Lunar \& Planetary Laboratory, University of Arizona, 1629 E. University Blvd., Tucson, AZ, USA}
\altaffiltext{6}{Geological and Planetary Sciences, California Institute of Technology, Pasadena, CA, USA}
\altaffiltext{7}{The University of Texas at Austin, Department of Astronomy, 2515 Speedway C1400, Austin, TX, USA}
\altaffiltext{8}{Department of Astronomy, California Institute of Technology, Pasadena, CA, USA}
\altaffiltext{9}{Astronomy Department, University of California, Berkeley, CA, USA}
\altaffiltext{10}{NASA Goddard}
\altaffiltext{11}{MIT Kavli Institute for Astrophysics and Space Research, 77 Massachusetts Ave, 37-241, Cambridge, MA, USA}
\altaffiltext{12}{UCO/Lick Observatory, Department of Astronomy \& Astrophysics, University of California, Santa Cruz, CA, USA} 
\altaffiltext{13}{Sagan Fellow}
\altaffiltext{14}{Hubble Fellow}
\altaffiltext{15}{Department of Terrestrial Magnetism, Carnegie Institute of Washington, Washington, DC, USA}
\altaffiltext{16}{Harvard-Smithsonian Center for Astrophysics, 60 Garden Street, Cambridge, MA, USA}
\altaffiltext{17}{Department of Astronomy, Yale University, 260 Whitney Avenue, New Haven, CT, USA}
\altaffiltext{18}{NASA Ames Research Center, Moffett Field, CA, USA}
\altaffiltext{19}{SETI Institute, 189 Bernardo Ave, Suite 200, Mountain View, CA, USA}
\altaffiltext{20}{SUPA, Institute for Astronomy, University of Edinburgh, Royal Observatory, Blackford Hill, Edinburgh, EH93HJ, UK}
\altaffiltext{21}{Observatoire Astronomique de l'Universit{\'e} de Gen{\'e}ve, 51 Chemin des Maillettes, 1290 Versoix, Switzerland}
\altaffiltext{22}{Centre for Exoplanet Science, SUPA, School of Physics \& Astronomy, University of St Andrews, St Andrews, KY16 9SS, UK}
\altaffiltext{23}{INAF, IASF Milano, Via E. Bassini 15, 20133 Milano, Italy}
\altaffiltext{24}{INAF-FGG, Telescopio Nazionale Galileo, La Palma, Spain}
\altaffiltext{25}{INAF, Osservatorio Astronomico di Palermo, Palermo, Italy}
\altaffiltext{26}{Dipartimento di Fisica e Astronomia, Universit{\'a} di Padova, Vicolo dell'Osservatorio 3, I-35122 Padova, Italy}
\altaffiltext{27}{INAF, Osservatorio Astronomico di Padova, Vicolo dell'Osservatorio 5, I-35122 Padova, Italy}
\altaffiltext{28}{Astrophysics Research Centre, School of Mathematics and Physics, Queen's University, Belfast BT7 1NN, UK}
\altaffiltext{29}{Centre for Star and Planet Formation, Natural History Museum of Denmark \& Niels Bohr Institute, University of Copenhagen, {\O}ster Voldgade 5-7, DK-1350 Copenhagen K, Denmark}
\altaffiltext{30}{Astronomy \& Astrophysics Department, Pennsylvania State University, University Park, PA 16802}
\altaffiltext{31}{Max-Planck-Institute for Astronomy, K{\"o}nigstuhl 17, 69117 Heidelberg, Germany}
%\altaffiltext{32}{Jet Propulsion Laboratory, California Institute of Technology, 4800 Oak Grove Drive, Pasadena, CA 91107, USA}
\altaffiltext{33}{Department of Astronomy and Astrophysics, University of Chicago, 5640 S. Ellis Ave, Chicago, IL 60637, USA}
\altaffiltext{34}{Department of Physics \& Astronomy, University of California Los Angeles, Los Angeles, CA 90095}
\altaffiltext{35}{INAF, Osservatorio Astrofisico di Torino, Via Osservatorio 20, 10025 Pino Torinese, Italy}
\altaffiltext{36}{Institut de Recherche sur les Exoplan{\`e}tes, Universit{\'e} de Montr{\'e}al, Montr{\'e}al, QC, Canada}
\altaffiltext{37}{Trottier Fellow}
%\altaffiltext{9}{NSF Graduate Student Fellow}
%\altaffiltext{7}{NSERC Postgraduate Research Fellow}
%\altaffiltext{13}{Lunar and Planetary Laboratory, University of Arizona, Tucson, AZ 85721 USA}
%\altafilltext{c}{Max Planck Institut fuer Astronomie, Koenigstuhl 17, Heidelberg, D-69117, Germany}

\begin{abstract}

HD~3167 is a bright ($V=8.9$), nearby K0 star observed by the NASA \emph{K2} mission (EPIC 220383386), hosting two small, short-period transiting planets. Here we present the results of a multi-site, multi-instrument radial velocity campaign to characterize the HD~3167 system. The masses of the transiting planets are \pmassberr\ M$_{\oplus}$ for HD~3167~b, a hot super-Earth with a likely rocky composition ($\rho_b$=\pdensberr~g~cm$^{-3}$), and \pmasscerr\ M$_{\oplus}$ for HD~3167~c, a warm sub-Neptune with a likely substantial volatile complement ($\rho_c$=\pdenscerr~g~cm$^{-3}$). We explore the possibility of atmospheric composition analysis and determine that planet c is amenable to transmission spectroscopy measurements, and planet b is a potential thermal emission target. We detect a third, non-transiting planet, HD 3167~d, with a period of \periodd\ d (between planets b and c) and a minimum mass of \pmassderr\ M$_{\oplus}$. We are able to constrain the mutual inclination of planet d with planets b and c: we rule out mutual inclinations below 1.3 degrees as we do not observe transits of planet d. From 1.3--40 degrees, there are  viewing geometries invoking special nodal configurations which result in planet d not transiting some fraction of the time. From 40--60 degrees, Kozai-Lidov oscillations increase the system's instability, but it can remain stable for up to 100Myr. Above 60 degrees, the system is unstable. HD~3167 promises to be a fruitful system for further study and a preview of the many exciting systems expected from the upcoming NASA \emph{TESS} mission.

\end{abstract}




\section{Introduction} 
\label{sec:intro}

One of the most interesting results of the previous decades of exoplanet discovery is the diversity in both the types of planets being discovered, and the types of planetary systems. In particular, the NASA \kepler\ mission \citep{Borucki10,Koch10} has revealed a large population of planets with sizes in between the radii of Earth and Neptune \citep[1--4R\earthsym; ][]{Howard2012,Fressin2013,Petigura2013,Burke2015}, a size range in which we have no examples in the Solar System. This presents an opportunity to map out the bulk composition of exoplanets as a function of their radius, and identify the size (or range of sizes) at which they transition from rocky (Earth-like) to volatile-rich (Neptune-like) compositions \citep{Weiss2014,Rogers2015,Wolfgang2015}. However, these relatively small planets produce correspondingly small radial velocity (RV) signals, which makes measuring their masses (and therefore bulk density) an expensive exercise. Therefore, the only feasible small exoplanets for characterization are those which orbit bright stars. The median apparent magnitude of the exoplanets discovered by \kepler\ in its original mission is 14.5 in the \kepler\ bandpass (400--900nm), and there are only seven planets in the 1--4R\earthsym\ range around stars brighter than 10th magnitude. Several of these, including Kepler-93b \citep{Dressing15} and Kepler-68b \citep{Marcy14} have been well-studied, and considerable effort has been expended on some fainter targets \citep[e.g. Kepler-78b, ][]{Grunblatt15}, but for robust investigation of the potential mass transition region, more data are required. One ground-based transit survey, the MEarth survey \citep{Berta-Thompson13} has provided two of these planets orbiting M dwarfs, around which these small planets provide a relatively large transit signal: GJ 1132b \citep{Berta-Thompson15} and GJ 1214b \citep{Charbonneau09}. However, the majority of ground-based transit surveys are limited in discovery space to larger planets. The discovery of the transiting nature of several radial-velocity planets, by selection orbiting bright stars, helps to fill out the sample, including HD 97658 b \citep{Howard2011,Dragomir13} and HD 219134 b \citep{Motalebi15}. Recently, the resurrection of the crippled NASA \kepler\ telescope as the \emph{K2} mission \citep{Howell14} has provided the community with a preview of the wide-field, shallow survey of bright stars that the NASA \emph{TESS} mission will complete \citep{Ricker2014}, focusing on targets which are highly amenable to further characterization. The discoveries by the \emph{K2} mission in this exoplanet size regime include three bright, nearby multi-planet systems: K2-3 b, c, and d \citep[$K=8.6$, ][]{Crossfield2015}, HIP~41378 b, c, and d \citep[$K=7.7$, ][]{Vanderburg2016a}, and HD~3167 b and~c \citep[$K=7.1$, ][]{Vanderburg2016b}. 

The bright targets discovered by \emph{K2} and \emph{TESS} will also provide some of the best targets for atmospheric characterization with NASA's \emph{James Webb Space Telescope (JWST)} \citep{Beichman2014,Greene2016}. Given the expected launch date for \emph{JWST} of 2018 October, %and the likely unavailability of TESS planet candidates until after the Cycle 1 Guest Observer proposals are due (2018 February), 
the aforementioned K2 discoveries are providing a timely supply of interesting, feasible observations for both Early Release Science and Cycle 1 observations. Measuring the masses of the planets is a key ingredient in interpreting the results of \emph{JWST} transmission and emission spectroscopy \citep{Benneke2012,Benneke2013}. 

%Measuring masses also important for constraining formation and evolution theories.

Here we present the results of a multi-instrument, multi-site campaign to characterize the masses of the planets in the HD 3167 system. The paper is organized as follows: in Section \ref{sec:transit} we describe the light curve and radial velocity data acquisition and analysis. In Section \ref{sec:pars} we describe the derived system parameters, including the likely composition. In Section \ref{sec:jwst} we examine the prospects for atmospheric characterization of the HD~3167 system, and finally in Section \ref{sec:dynamics} we analyze the architecture and dynamical stability of the HD~3167 system.

\section{Observations and Data Analysis}

\subsection{Transit detection}
\label{sec:transit}

\begin{figure*}[ht!]
%\figurenum{1}
%\plotone{lightcurves.pdf}
\includegraphics[width=\textwidth]{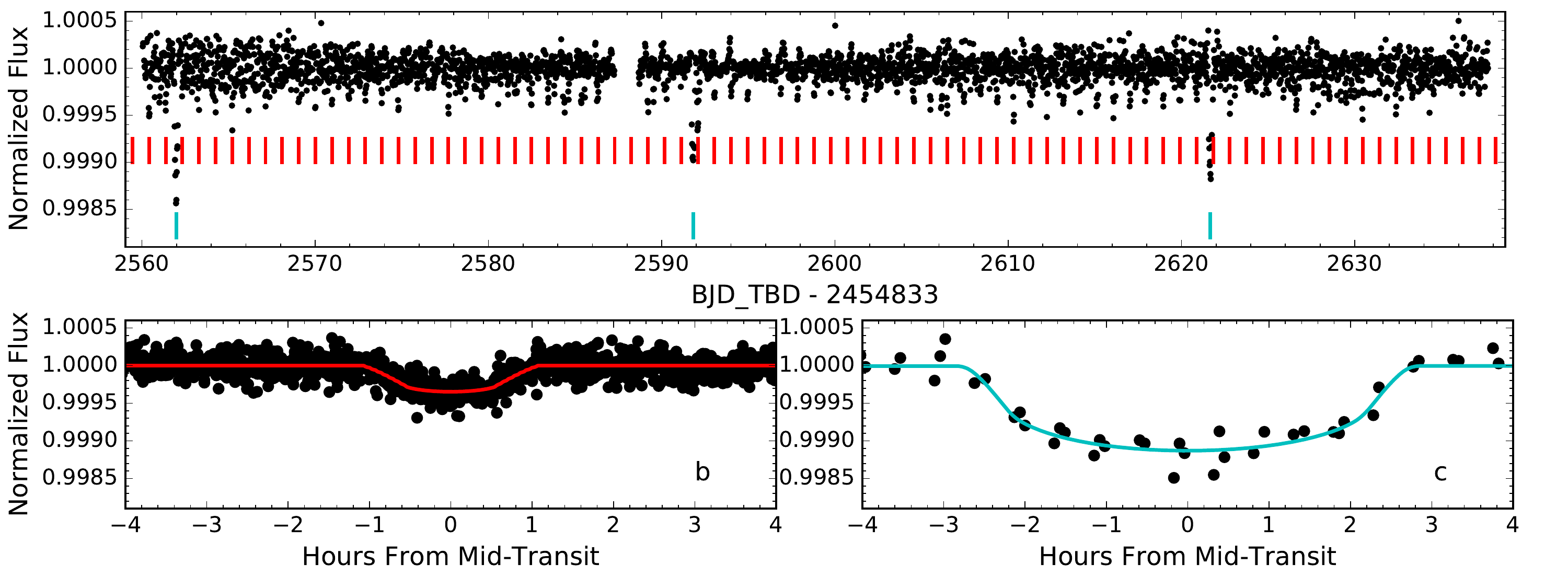}
\caption{The top panel shows the detrended K2 photometry for HD 3167. Transit of planet b are marked in red, and transits of planet c in cyan. The bottom panels show the phase-folded K2 photometry for planets b (left) and c (right). The best fit transit model, described in Section \ref{sec:pars}, is over-plotted in red for planet b, and cyan for planet c.\label{fig:lightcurve}}
\end{figure*}

The NASA \emph{K2} mission uses the \kepler\ spacecraft to observe a series of fields, called campaigns, around the ecliptic plane. Near-continuous, high-precision photometry is obtained on 10,000--20,000 targets per campaign, most targets having 30-minute integrations. Campaign 8 (C8) awas observed for 80 days from 2016 January 04 to 2016 March 23. The calibrated pixels were downloaded from the Mikulski Archive for Space Telescopes (MAST) and processed in the same fashion as \citet{Crossfield2015}. In brief, following the methods of \citet{Vanderburg2014} and \citet{Vanderburg2014a}, the photometry is divided into six roughly equal segments, and each is decorrelated against the location of the photocenter of the light using a 1D Gaussian process. The major systematic in the photocenter location is the roll of the spacecraft around the telescope foresight, which is corrected approximately every six hours. By switching antennae at the start of Campaign 8, the magnitude of the roll was reduced significantly from that seen in Campaign 7, resulting in overall higher quality light curves with higher precision\footnote{\url{http://keplerscience.arc.nasa.gov/k2-data-release-notes.html}}. One of the targets observed in Campaign 8 was HD~3167, a bright (V=8.9, K=7.0), nearby (\distance\ pc), K0 dwarf star, also designated as \epicno. The detrended photometry is shown in the top panel of Figure \ref{fig:lightcurve}.

Three transits of a long-period, relatively deep ($\sim$1 mmag) planet candidate were first detected in a by-eye search of the brightest targets in C8, marked in cyan in the top panel of Figure \ref{fig:lightcurve}. On closer inspection, shallower transits at a significantly shorter period were also detected, marked in red. Using the TERRA algorithm \citep{Petigura2013}, two signals were found with periods of 0.959609 days (shown in the bottom left panel of Figure \ref{fig:lightcurve}) and 29.8479 days (shown in the bottom right panel), with transit depths of \depthb\ ppm and \depthc\ ppm, respectively. These signals were subsequently reported by \citet{Vanderburg2016b} as HD~3167~b and c respectively. After removal of those signals, no additional transiting signals were found with an SNR above $5\sigma$, corresponding to $\sim$0.8~R$_\oplus$. 

In addition, we performed several tests of the photometry to rule out obvious false positive scenarios prior to acquiring expensive, high-precision radial velocity measurements. These included an adaptation of the model-shift uniqueness test, originally designed for \kepler\ data and described in Section 3.2.3 of \citet{Coughlin2016}. In brief, the test searches for other significant transit-like events in the light curve when phased to the period of the putative planet signal: false positives will often show multiple significant events across all phases due to the higher levels of correlated noise. Both planets b and c passed the model-shift uniqueness test. We also included an adaptation of the Locality Preserving Projections (LPP) test, described in \citet{Thompson2015} for \kepler\ data, which uses dimensionality reduction and $k$-nearest neighbours to measure how similar a putative signal is to a planetary transit signal. Both planets b and c also passed the LPP test. 

\begin{figure}[b]
\centering
%\figurenum{2}
%\includegraphics[width=\columnwidth]{centroids_first_planet_b.pdf}
\includegraphics[width=\columnwidth]{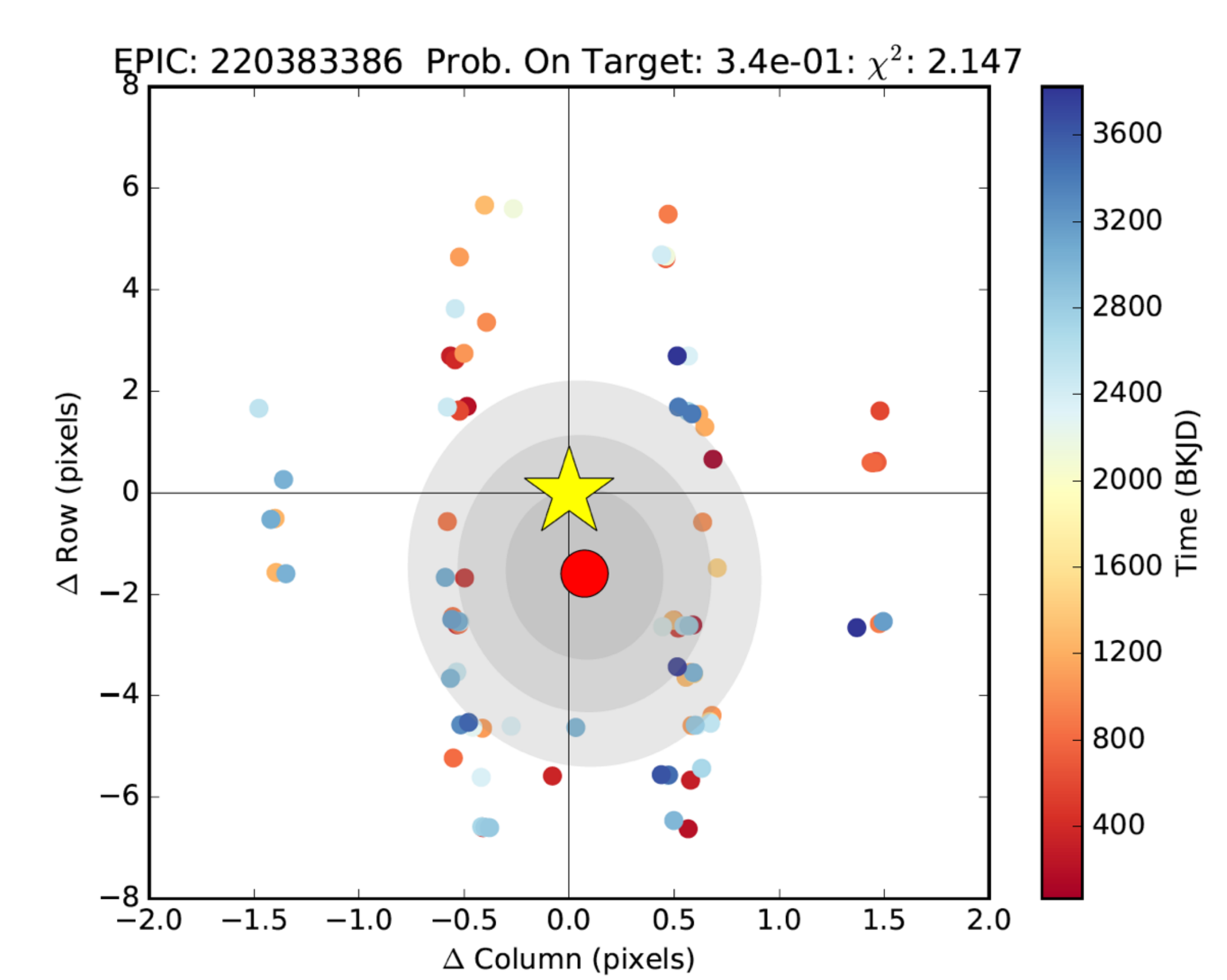}

\caption{The locations of the measured photocenters of light during the transits of HD~3167~b. There is a larger scatter in the row direction because HD~3167 is highly saturated. The locations are consistent with HD~3167 being the source of the transit signal. BKJD$=$BJD$-2454833.0.$\label{fig:centroids}}
\end{figure}

Finally, we examined the photocenters of light during the transits of planet b: significant motion of the photocenter of light away from the location of the putative host star during transit is a powerful technique for detecting false positive events due to background eclipsing binaries. For the original \kepler\ mission, the spacecraft pointing stability was so high that this method could be used to identify false positives lying well within the same pixel as the target star. We adapt the difference imaging technique used in Kepler \citep{Bryson2013} to \emph{K2}. Due to the strong roll motion in \emph{K2}, there is a large change in the light distribution between two consecutive cadences. even in the absence of a transit. Instead, for each in-transit cadence, we look for out-of-transit cadences at the same roll angle and and separated by exactly one thruster firing event. The roll angle is measured by the same technique as \citet{Vanderburg2014}. By requiring the out-of-transit cadences to be close in time, we minimize of the impact of motion perpendicular to the roll axis due to, e.g, differential velocity abberation. However, the \emph{K2} roll motion is not exactly repeatable, and not all in-transit cadences have out-of-transit cadences that meet our requirement both before and after the transit. For cadences that do, we fit the PRF model of \citet{Bryson2010} to the in-transit and difference image, and compute the shift in the photocenter. We average over all cadences for which a difference can be computed, and calculate the probability that the observed distribution of offsets is consistent with the hypothesis that the location of the transit is consistent with the location of the target star. For simplicity, the distribution of offsets is assumed to be Gaussian in both row and column. At 9th magnitude, HD~3167 is highly saturated, resulting in large scatter in the row direction due to the bleed of saturated pixels, and the distribution is highly non-Gaussian. Nevertheless, Figure \ref{fig:centroids} shows no strong evidence that the source of the transit for planet b is offset from the target. There are only three transits of HD~3167~c, and one of those gives a poor fit to the photocenter location, so we do not perform the photocenter analysis on this planet.

\subsection{Stellar characterization}
\label{sec:star}

\begin{figure}[t]
%\figurenum{3}
%\plotone{HD3167_mgb_seg_fit2.pdf}
\includegraphics[width=\columnwidth]{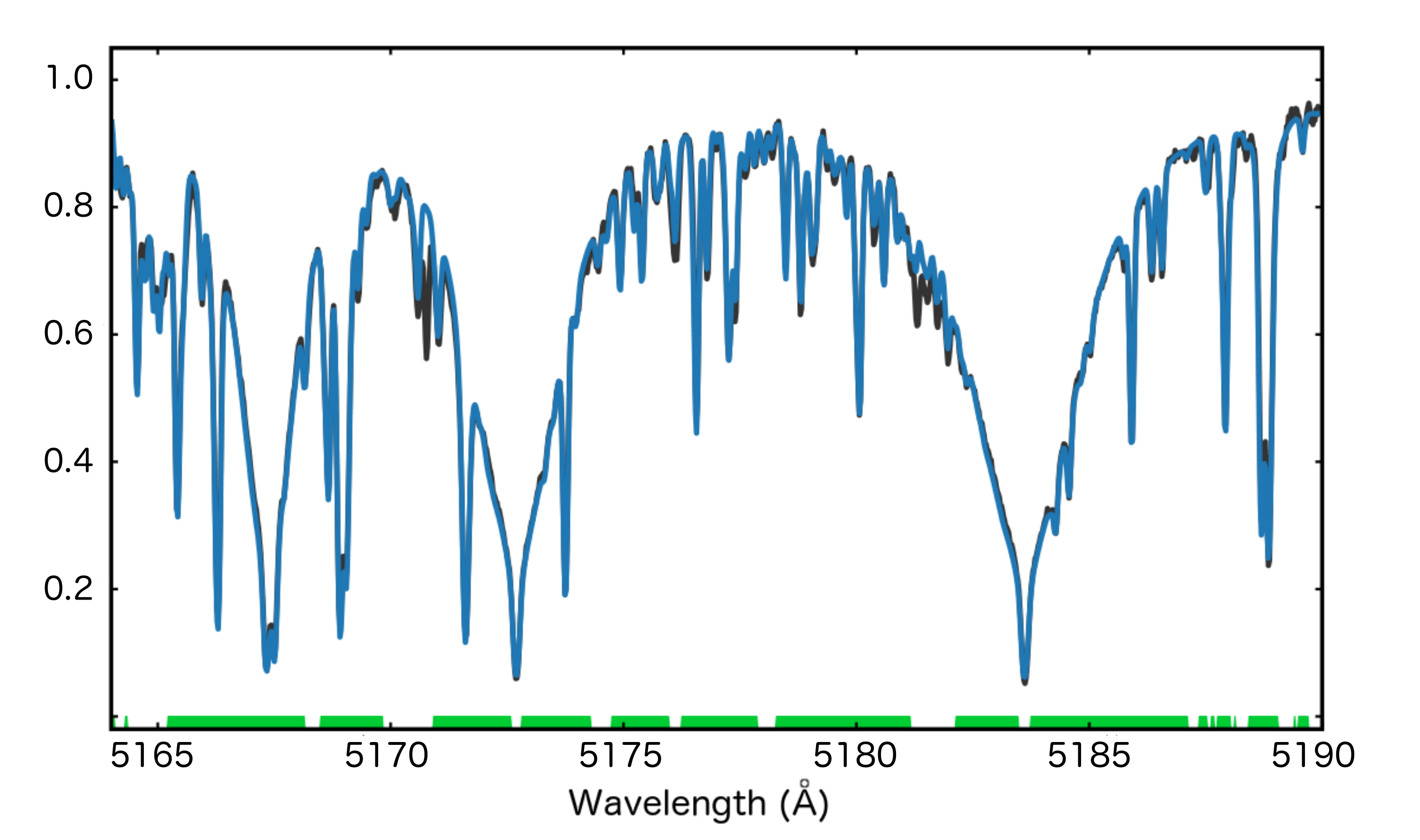}
\caption{Final model fit to one of the Keck/HIRES template spectra used to derive the stellar parameters in the region of the Mg b triplet.  The black line is the observation, light blue is the model, and the green line at the bottom indicates the regions used in the fitting.  There were 350~\AA\ used in the full fit in regions between 5164~\AA\ and 7800~\AA.\label{fig:spectrum}}
\end{figure}

To determine the stellar parameters of the host star, we obtained three %iodine-free %template
spectra of HD~3167 using Keck-HIRES with a S/N of $\sim 260$ at 6000~\AA, without using the iodine cell as is typical of the precision radial velocity observations; Figure \ref{fig:spectrum} shows a segment of a spectrum in the region of the Mg b triplet. We derived the stellar properties using the spectral forward-modeling procedure and line list of \citet{Brewer2016}. We first fit for $T_{\mathrm{eff}}$, $\log g$, [M/H], and Doppler broadening using a scaled solar abundance pattern except for the alpha elements calcium, silicon, and titanium. We then fixed the stellar parameters and solved for the abundances of 15 elements.  Finally, we repeated the process using this new abundance pattern. The results from fitting the three different spectra were nearly identical for all parameters.  We then apply the empirical corrections from \citet{Brewer2016} to obtain the final parameters, summarized in Table \ref{tab:stellarpars}.

The analysis procedure has been shown to recover gravities consistent with those of asteroseismology with an RMS scatter of 0.05 dex \citep{Brewer2015} and we adopt this as the uncertainty in $\log g$. \citet{Brewer2016} shows that there is a 39~K offset with temperatures derived from well-measured angular diameters. We add this in quadrature to their 25~K statistical uncertainties for a total uncertainty of 46~K. The statistical uncertainty in the [Fe/H] measurement is only 0.01~dex but the empirical correction at this temperature is 0.09~dex. We adopt half of the offset, 0.05~dex, as our uncertainty in [Fe/H]. Finally, we compare the results of the analysis to those given by the Stellar Parameter Classification tool \citep[SPC;][]{Buchhave2012,Buchhave2014} and SpecMatch \citep{Petigura2015} and find they agree to within 1$\sigma$. Following the procedure in \citet{Crossfield2016}, we use the free and open source {\texttt isochrones} Python package \citep{Morton2015} and the Dartmouth stellar evolution models \citep{Dotter2008} to estimate the stellar radius and mass given in Table \ref{tab:stellarpars}. The resulting stellar density is consistent with values derived in the transit analyses. HD~3167 was not included in Gaia Data Release 1 \citep{Gaia2016}, possibly due to the incompleteness at the bright end or the poorer coverage along the ecliptic, where the \emph{K2} mission observes by necessity. However future Gaia releases should produce a precise distance and allow for stronger constraints on the stellar parameters. From the HIPPARCOS parallax \citep{Perryman1997}, and following the same procedure as \citet{Brewer2016}, we derive an age for HD~3167 of 7.8$\pm4.3$~Gyr.

The \emph{K2} data show some longer term variability that may be caused by stellar rotation (see Fig. 1 of Vanderburg et al. 2016b). Examining the auto-correlation function of the light curve reveals a broad peak from 20--35 days, with a maximum at 27.2 days. The rotational velocity of 1.7$\pm$1.1 km s$^{-1}$ is fairly poorly constrained, and allows a range of rotational periods from 10--40 days. These values are broadly consistent with the expected value for a field K-dwarf \citep[see, e.g.][]{Newton2016}. We examine the correlations between stellar activity indicators and the measured radial velocities in Section \ref{sec:search}.

\begin{figure}[t]
%\figurenum{4}
\includegraphics[width=\columnwidth]{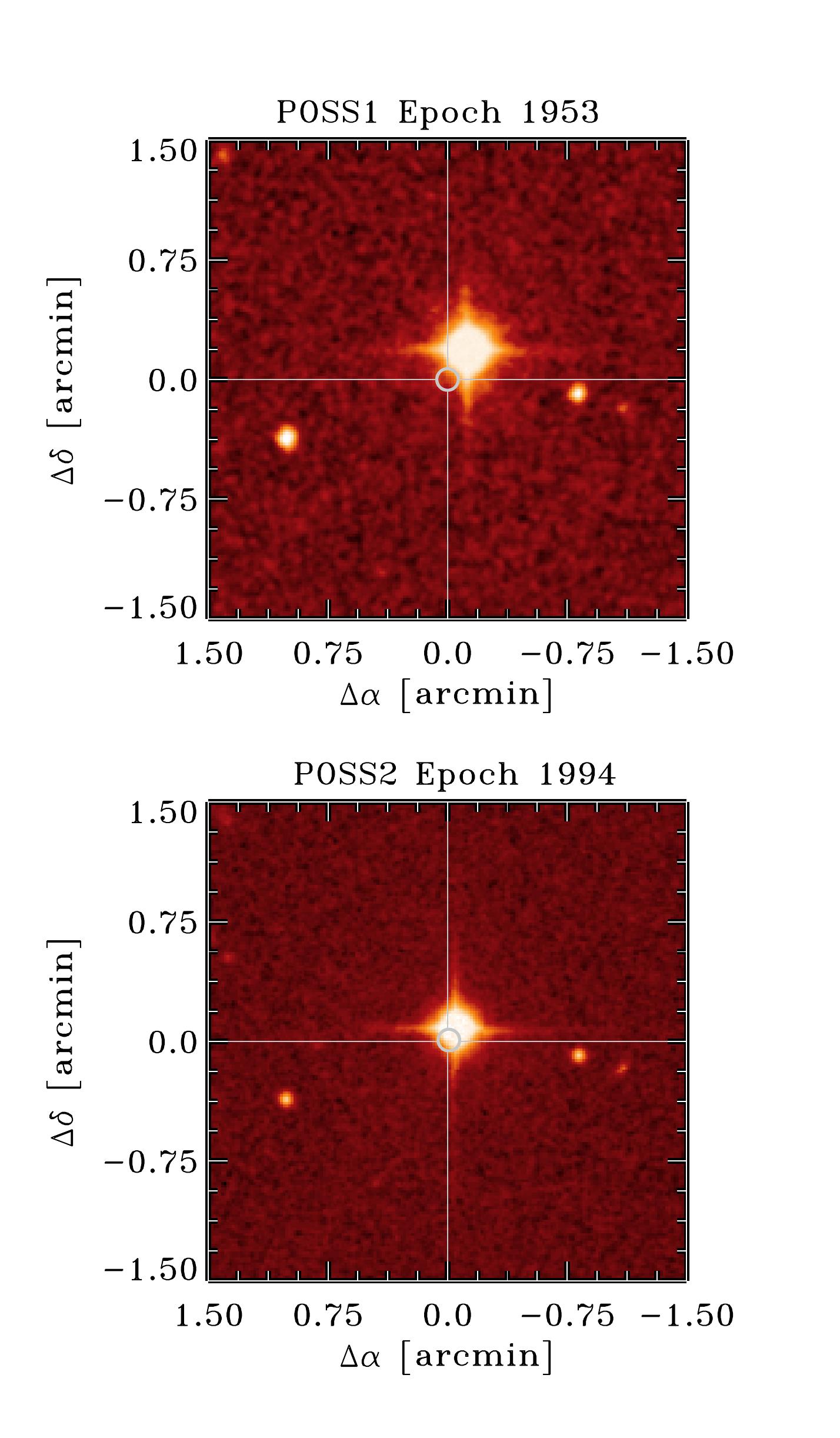}
\caption{POSS1 red plates observed in 1953 (top panel) and POSS2 red plates observed in 1994 (bottom panel). The circle shows the location of HD~3167 at the 2016 position of the star. Between 1953 and 1994, HD~3167 moved by $\sim 8$ arcsec, which can be clearly seen in the DSS images. The POSS1 plate rules out a background star coincident with the current location of HD3167 to ${\rm{\Delta }}R \approx 5$ mag. \label{fig:propermotion}}
\end{figure}

\subsubsection{Proper motion}
The proper motion of HD~3167 is quite large \citep[107 mas/yr in right ascension and -173 mas/yr in declination; ][]{Huber2016}. In the 63 years since the 1953 Palomar Observatory Sky Survey (POSS) images, HD 3167 has moved more than 12.5$^{\prime\prime}$, enabling us to utilize archival POSS data to search for background stars that are now, in 2016, hidden by HD 3167. Using the 1953 POSS data, shown in the top panel of Figure \ref{fig:propermotion}, we find no evidence of a background star at the current postion of HD 3167 to a differential magnitude of $\sim$5 magnitudes, shown in the bottom panel of Figure \ref{fig:propermotion}. Because HD 3167 is saturated in the POSS images, this sensitivity was estimated by placing fake sources at the epoch 2016 position of HD 3167 in the epoch 1953 image and estimating the 5$\sigma$ threshold for detection. The photometric scale of the image (and hence, the magnitudes of the injected test stars) was set using the star located 1$^\prime$ to the southeast of HD 3167, which has an optical magnitude of approximately B$=$15.5. This analysis does not rule out the most extreme background eclipsing binaries (a 50\% eclipsing binary would produce a 1~mmag transit at a differential magnitude of 6.8 magnitudes), but was sufficient for us to instigate the high-precision radial velocity campaign. 

\subsubsection{Adaptive Optics}
We obtained near-infrared adaptive optics images of HD~3167 at Keck Observatory on the night of 2016 July 14 UT. Observations were obtained with the $1024 \times 1024$ NIRC2 array and the natural guide star system; the target star was bright enough to be used as the guide star. The data were acquired using the narrow-band Br-$\gamma$ filter using the narrow camera field of view with a pixel scale of 9.942 mas/pixel. The Br-$\gamma$ filter has a narrower bandwidth (2.13--2.18 $\mu m$), but a similar central wavelength (2.15 $\mu m$) compared the Ks filter (1.95-2.34 $\mu m$; 2.15 $\mu m$) and allows us to observe HD 3167 without saturation. A 3-point dither pattern was utilized to avoid the noisier lower left quadrant of the NIRC2 array. The 3-point dither pattern was observed with 10 coadds and a 0.726 second integration time per coadd for a total on-source exposure time of 65 s.

\begin{figure}[t]
%\figurenum{5}
\includegraphics[width=\columnwidth]{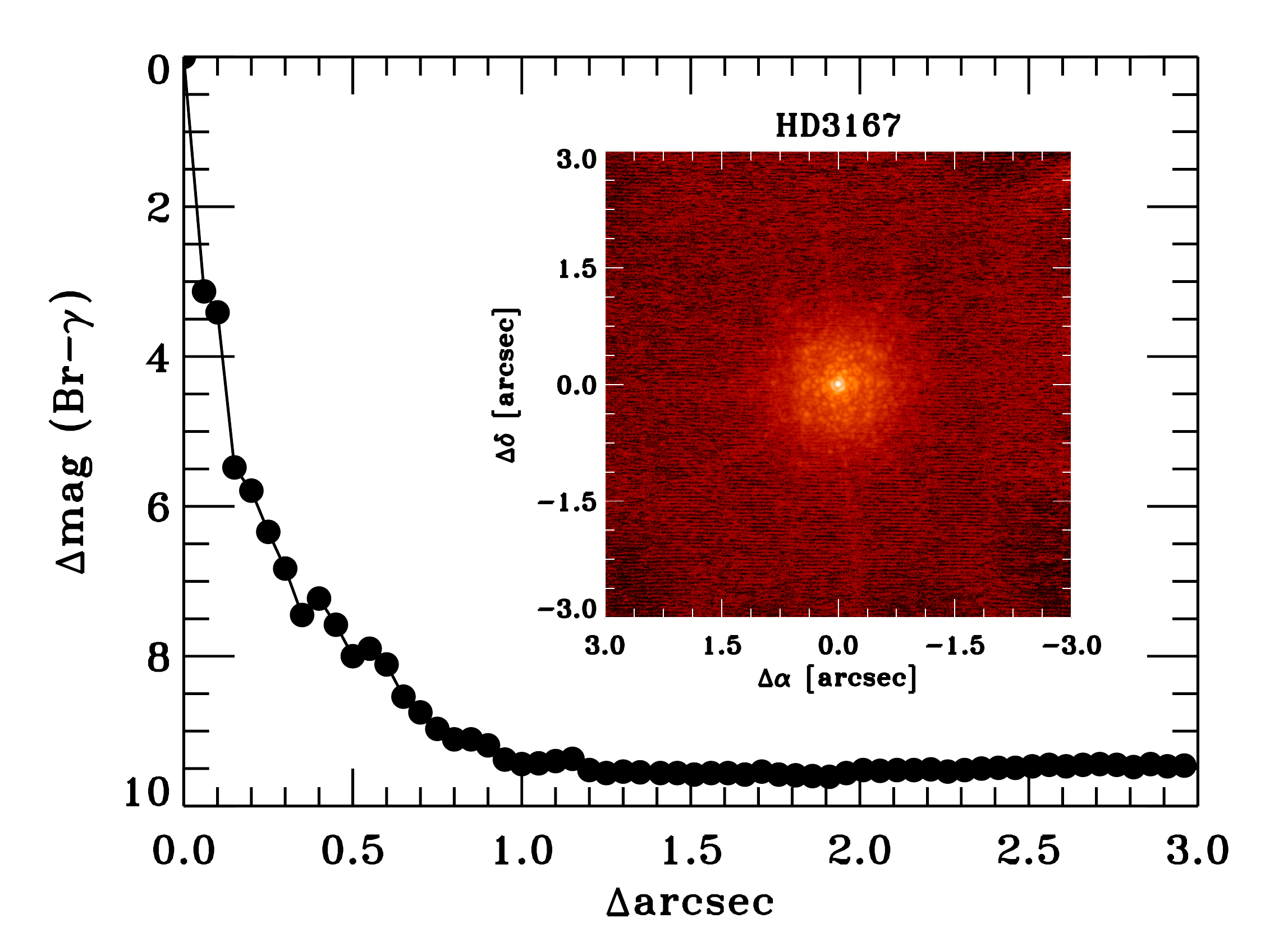}
\caption{Keck Observatory NIRC2 K-band image and the associated contrast curve. No stars with contrasts ${\rm{\Delta }}K< 3.4\;\mathrm{are}$ detected with separations $>0.1$ arcsec and ${\rm{\Delta }}K< 8.0$ with separations $>0.5$ arcsec. \label{fig:aolimits}}
\end{figure}

HD~3167 was measured with a resolution of 0.050$^{\prime\prime}$ (FWHM). No other stars were detected within 4$^{\prime\prime}$ of HD~3167. In the Br-$\gamma$ filter, the data are sensitive to stars that have K-band contrast of $\Delta$K = 3.4 mag at a separation of 0.1$^{\prime\prime}$ and $\Delta$K$=$8.0 mag at 0.5$^{\prime\prime}$ from the central star. We estimate the sensitivities by injecting fake sources with a signal-to-noise ratio of 5 into the final combined images at distances of $N\times$ FWHM from the central source, where $N$ is an integer. The 5$\sigma$ sensitivities, as a function of radius from the star, are shown in Figure \ref{fig:aolimits}. Beyond 4$^{\prime\prime}$, there are no additional stars visible in 2MASS out to a radius of $\sim$20$^{\prime\prime}$.

\begin{deluxetable}{lll}
\tablecaption{HD~3167 stellar parameters \label{tab:stellarpars}}
\tablehead{ 
    \colhead{Parameter}    & \colhead{Value}  & \colhead{Units}\\
}
%\colnumbers
\startdata
RA & 00:34:57.52 & hh:mm:ss\\
Dec  & $+$04:22:53.3 & dd:mm:ss \\
EPIC ID & EPIC 220383386 & \\
2MASS ID & 2MASS J00345752$+$0422531 & \\
V & 8.941$\pm$0.015 & mag\\
K & 7.066$\pm$0.020 & mag\\
Spectral Type & K0 V & \\
$T_{\rm{eff}}$ & 5261$\pm$60 & K\\
log $g$  & 4.47$\pm$0.05 & log$_{10}$(cm s$^{-2}$)\\
R$_{\star}$ & 0.86$\pm$0.04 & R\sunsym \\
M$_{\star}$  & 0.86$\pm$0.03 & M\sunsym \\
$\rho_{\star}$\tablenotemark{a} & 1.902$\pm$0.092 & g cm$^{-3}$ \\
$\rho_{\star,b}$\tablenotemark{b} & 1.40$^{+0.52}_{-0.79}$ & g cm$^{-3}$ \\
$\rho_{\star,c}$\tablenotemark{c} & 1.39$^{+0.65}_{-0.94}$ & g cm$^{-3}$  \\
Distance & 45.8$\pm$2.2\tablenotemark{d} & pc\\
{[}Fe/H{]} & 0.04$\pm$0.05 & \\
$v$ sin $i$ & 1.7$\pm$1.1 & km s$^{-1}$\\
log R'HK & -5.04 & \\
\enddata
\tablenotetext{a}{Spectroscopically derived}
\tablenotetext{b}{Derived from transit light curve fit to planet $b$}
\tablenotetext{c}{Derived from transit light curve fit to planet $c$}
\tablenotetext{d}{\citet{vanLeeuwen2007}}
%\tablenotetext{e}{Time of inferior conjunction}

%\tablecomments{Note that {\tt \string \colnumbers} does not work with the 
%vertical line alignment token. If you want vertical lines in the headers you
%can not use this command at this time.}
\end{deluxetable}

\clearpage

\subsubsection{Radial velocity measurements}

After the identification and validation of the two transiting planet signals, a high-cadence observing campaign was rapidly launched in order to obtain mass measurements while C8 was still visible. The final data set includes observations obtained with Keck/HIRES, APF/Levy, and HARPS-N, described below. The full set of radial velocity measurements is given in Table \ref{tab:rv}.

Our observational setup for both Keck/HIRES and the APF/Levy was essentially identical to those described in \citet{Fulton2016} and \citet{Burt2014}. We collected a total of 60 RV measurements using Keck/HIRES \citep{Vogt1994}, and 116 measurements using the Levy Spectrograph on the Automated Planet Finder \citep[APF,][]{Vogt2014,Radovan2014} at Lick Observatory between 2016 July 7 and 2016 December 2. For all of the Keck/HIRES measurements we collected three consecutive exposures in order to mitigate the affects of stellar oscillations \citep{Dumusque2011}. The three measurements were then binned together before a jitter term, which includes a contribution from stellar jitter, is added in quadrature during the modeling process (see Section \ref{sec:pars}); this technique was not necessary at APF due to the smaller telescope aperture and longer exposure times. Whenever possible we observed HD 3167 two times during a single night with maximum temporal separation to improve phase coverage for HD 3167 b.

Each Doppler spectrum was taken through a cell of gaseous iodine that imprints a dense forest of molecular absorption lines onto the stellar spectrum and serves as both a wavelength and point spread function (PSF) reference. The slits chosen provided spectral resolving power of $R\sim$70,000 and $R\sim$100,000 for Keck and APF respectively. A series of iodine-free spectra were also collected using a narrower slit on both instruments ($R\sim$85,000/120,000 for Keck/APF). These spectra were deconvolved with the instrumental PSF and used as models of the intrinsic stellar spectrum. We modeled each RV observation as the deconvolved intrinsic stellar spectrum shifted by a best-fit RV and multiplied by an ultra-high resolution iodine transmission spectrum. This is then convolved with an instrumental PSF, which is modeled as the sum of 13/15 Gaussians for Keck/APF \citep{Butler1996}. We reject measurements with SNR$<$45, mid-exposure times before or after 13-degree twilight, and measurements collected when the star was within 20 degrees of the moon. The rejected observations are not included in Table \ref{tab:rv}.

We also observed HD 3167 with the HARPS-N spectrograph \citep{Cosentino2012} located at the 3.58m Telescopio Nazionale Galileo (TNG) on the island of La Palma, Spain. HARPS-N is a stabilized spectrograph designed for precise radial velocity measurements. We observed HD 3167 76 times between 2016 July 7 (independently beginning the same night as the HIRES/APF campaign) and 2016 December 7, obtaining high resolution optical spectra with a spectral resolving power of $R = 115000$. Most of our observations consisted of 15 minute integrations, which yielded formal photon-limited Doppler uncertainties between 0.6 and 1.6 m/s. Similarly to the Keck/HIRES measurements, we typically observed HD 3167 two times per night, separated by a couple hours, in order to better sample the inner planet's orbit; on several occasions, we observed HD 3167 up to six times per night. We measured radial velocities by calculating a weighted cross-correlation function between the observed spectra and a binary mask \citep{Baranne1996,Pepe2002}.

\begin{deluxetable}{cccc}
\tablecaption{Radial Velocities. \label{tab:rv}}
%\tablewidth{245pt}
\tablehead{ 
    \colhead{HJD$_{\rm UTC}$}      & \colhead{RV\tablenotemark{1}}  & \colhead{Unc.\tablenotemark{2}} & \colhead{Inst.} \\
    \colhead{(-- 2440000)}  & \colhead{(\mse)}            & \colhead{(\mse)}       & \colhead{}
}
\startdata
 17580.116198 &     -3.355832 &  1.046724 &   HIRES \\
 17580.119335 &     -4.718952 &  0.969771 &   HIRES \\
 17580.122437 &     -4.812388 &  1.010754 &   HIRES \\
 17576.954515 &     -5.719372 &  1.449235 &     APF \\
 17576.976496 &     -9.079391 &  1.197070 &     APF \\
 17578.949488 &     -8.080386 &  1.427537 &     APF \\
 17673.539683 &  19528.630000 &  0.820000 &  HARPS-N \\
 17673.581963 &  19527.160000 &  0.790000 &  HARPS-N \\
 17673.625133 &  19524.580000 &  0.780000 &  HARPS-N \\
\enddata
\tablenotetext{}{(This table is available in its entirety in a machine-readable form in the online journal. A portion is shown here for guidance regarding its form and content.)}
\tablenotetext{1}{Zero point offsets between instruments have not been removed and must be fit as free parameters when analyzing this dataset}
\tablenotetext{2}{Stellar jitter has not been incorporated into the uncertainties.}
\vspace{10pt}
\end{deluxetable}

\begin{figure}[t]
\includegraphics[width=\columnwidth]{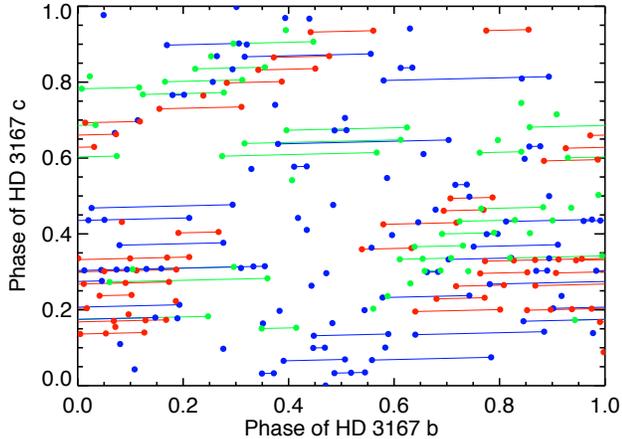}
\caption{The coverage of the phase combinations between planets b and c. The yellow diamonds are HARPS-N observations, the black open circles are HIRES observations, and the green points are APF observations. The solid lines connect observations obtained on the same night. HARPS-N and HIRES have only partial coverage of the phase combinations; APF has near-complete coverage. \label{fig:phases}}
\end{figure}

One factor that extended our radial velocity campaign was the aliasing between the $\sim$1-day orbital period of planet b and the $\sim$1-month orbital period of outer planet c. Given the restrictions on observing enforced by the diurnal cycle and the tendency for telescope time to be allocated approximately monthly around the full moon, it was difficult at any single longitudinal site to secure the required phase coverage to break the degeneracy between planets b and c. Figure \ref{fig:phases} shows, for each of the three telescopes, the b and c phase combinations of the observations. The HARPS-N observations, shown as yellow diamonds, represent the most precise measurements in our radial velocity sample, but have large bands of phase combinations that are un-sampled. Similarly, the HIRES measurements, shown as black open circles, do not cover the full range of phase combinations. Early analyses of the radial velocities from either of these sites individually led to degeneracies in the radial velocity semi-amplitudes, and therefore masses, of the b and c planets. The APF observations, shown as green points, for which there is the most regular access to the telescope, provide comprehensive coverage of the phase combinations of planets b and c. By combining the higher precision but limited phase coverage observations from HARPS-N and HIRES with the lower precision but broad phase coverage of APF we break the degeneracies and constrain the orbital solution as discussed below. %The solution will be further improved by adding precise data from another site, such as HARPS South, to fill in the phase coverage, and by continuing RV observations until the phase coverage is complete.

\section{System parameters}
\label{sec:pars}

\subsection{Transit analysis}

\begin{figure}[t]
\includegraphics[width=\columnwidth]{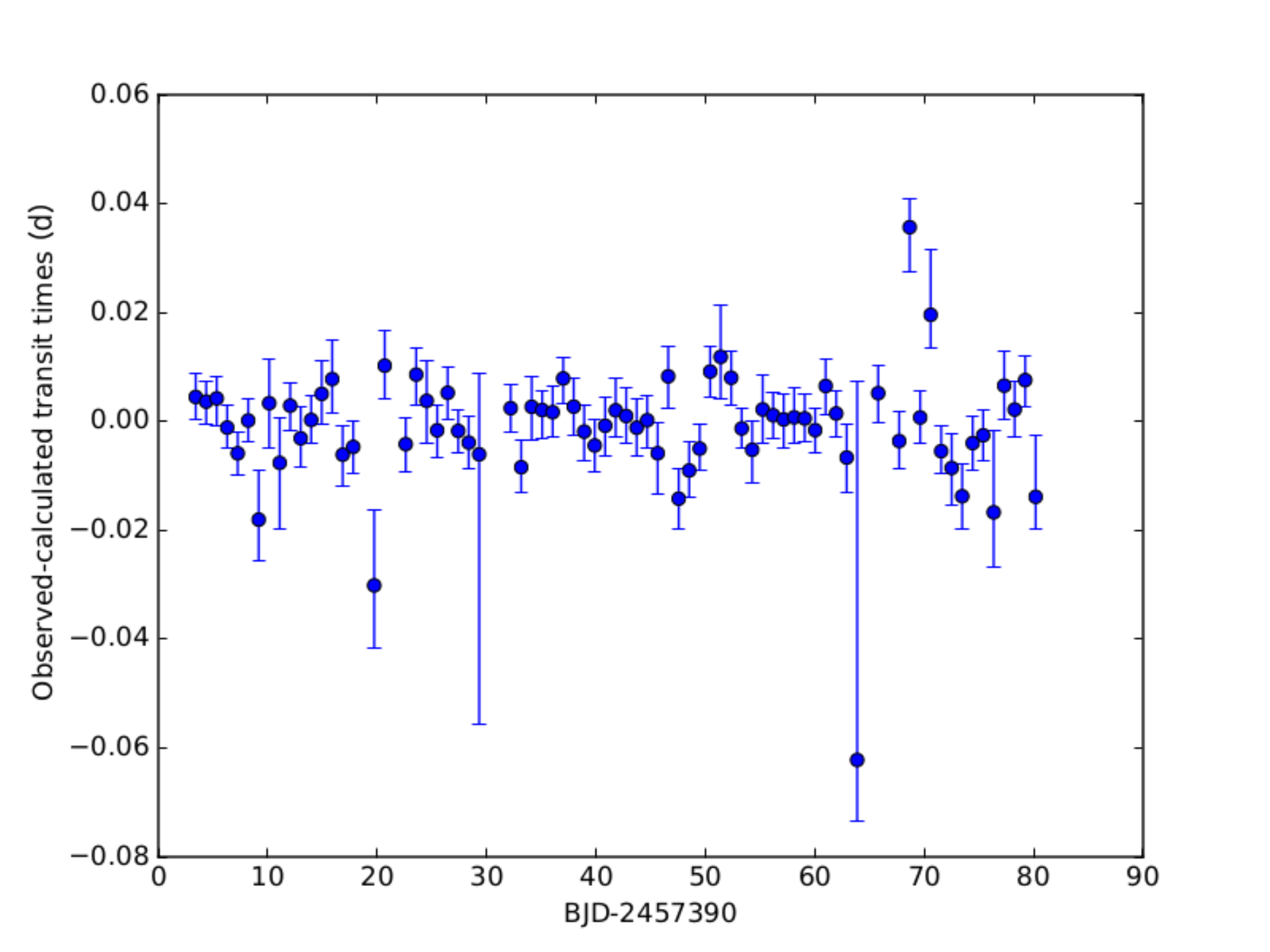}
\caption{The transit times of HD~3167 b in the K2 C8 light curve, compared to a linear ephemeris. At only 1.6 hours, the transit duration of planet b is short and poorly sampled by the 30-minute observation cadence. The average timing precision is $\sim$15 minutes.\label{fig:ttvs}}
\end{figure}

We analyzed the transit signals for planets b and c independently in our light curve, using the same modeling, fitting, and MCMC procedures as described in \citet{Crossfield2016}. As in that analysis, eccentricity was held to zero; for the radial velocity analysis described in Section \ref{sec:rvanalysis} we allowed the eccentricity of planet c to float. The results are shown in Table \ref{tab:pars} and are consistent with the parameters given by \citet{Vanderburg2016b} for planets b and c. We examine the transit times of planet b and find no evidence of variations above the level of $\sim$15 minutes, shown in Figure \ref{fig:ttvs}. Occasional outliers are present in the individually derived transit times, but we conclude that these are likely a result of the low cadence of the \kepler\ observations combined with a non-perfect detrending. We exclude cadences affected by spacecraft thruster firings prior to analysis. In addition, we apply the cosmic-ray detection algorithm for \emph{K2} photometry developed by \citet{Benneke2016}, but do not identify any cosmic-ray events as the source for the outliers in the transit timing. 

\subsection {Radial velocity analysis}
\label{sec:rvanalysis}

We analyzed the RV time-series using the publicly-available RV fitting package \texttt{RadVel}(Fulton \& Petigura, in prep.)\footnote{http://radvel.readthedocs.io/en/master/index.html}. \texttt{RadVel} is written in object-oriented Python and is designed to be highly extensible, flexible, and documented for easy adaptation to a variety of maximum-likelihood fitting and MCMC applications. The standard version of \texttt{RadVel} downloadable from GitHub\footnote{https://github.com/California-Planet-Search/radvel} includes a pipeline that is capable of modeling multi-planet, multi-instrument RV time-series utilizing a fast Keplerian equation solver written in C.

Our likelihood function for this analysis follows that of \citet{Sinukoff2016}:
\begin{equation}
\ln{\mathcal{L}} = - \sum_i{\left[ \frac{(v_i - v_m(t_i))^2}{2(\sigma_i^2 + \sigma_{\rm j}^2)} + \ln{\sqrt{2\pi(\sigma_i^2 + \sigma_{\rm j}^2)}}  \right]},
\end{equation}
where $v_i$ are the gamma-subtracted velocity measurements ($v_i = v_{i,\rm inst} - \gamma_{\rm inst}$, where $\gamma_{\rm inst}$ is an instrument-dependent term) with associated uncertainties $\sigma_i$, and $v_mk(t_i)$ is the Keplerian model at time $t_i$.

We first find the maximum-likelihood model using the Powell minimization technique \citep{Powell1964} then perturb the best-fit parameters by 1 part in $10^5$ to start 50 parallel MCMC chains. \texttt{RadVel} incorporates the affine-invariant sampler of the \texttt{emcee} package \citep{Foreman-Mackey2013}. The Gelman-Rubin \citep{Gelman2003} and $T_{z}$ statistics \citep{Ford2006} are checked in real-time during the MCMC exploration. The chains are deemed well-mixed and the MCMC is halted when the Gelman-Rubin statistic is within 3\% of unity and $T_{z}>1000$ for all free parameters. We chose to parameterize the Keplerian orbits using $\sqrt{e}\sin{\omega}$  and $\sqrt{e}\cos{\omega}$ instead of $e$ and $\omega$ in order to increase convergence speed. We assigned uniform priors to $\sqrt{e}\sin{\omega}$, $\sqrt{e}\cos{\omega}$, velocity semi-amplitudes ($K$), and the zero-point offsets ($\gamma$). The jitter terms for each instrument ($\sigma_{\rm{j}}$) are defined in Equation 2 of \citet{Fulton2015}, and serve to capture the stellar jitter and instrument systematics such that the reduced $\chi^2$ of the best-fit model is close to 1. The $\chi^2$ values in Table \ref{tab:comp} are reported without including the jitter terms, since including them would artificially reduce the final $\chi^2$ values. Gaussian priors were assigned to the ephemerides of the two transiting planets using the values reported in \citet{Vanderburg2016b}. We examine the fits for system architectures from 0--3 planets and choose the three-planet solution favored by the Bayesian information criterion (see Table \ref{tab:comp} for details). The median values and the 68\% credible intervals of the three-planet solution are reported in Table \ref{tab:params}. The best-fit three-planet Keplerian model is shown in Figure \ref{fig:rvmultipanel}.

\subsubsection{Search for a third planet}
\label{sec:search}

We search for additional planets in the RV data using the automated planet discovery pipeline described in \citet{Fulton2016} and \citet{Howard2016}. In brief, this pipeline utilizes a custom implementation of the two-dimensional Keplerian Lomb-Scargle periodogram \citep[2DKLS;][]{Otoole09}. Periodogram power is defined as a change in $\chi^2$ relative to a baseline $\chi^2$. For this particular search the baseline $\chi^2$ is derived from the best two-planet model fit. The periodogram, shown in the top panel of Figure \ref{fig:multi-peri}, demonstrates the change to the fit when adding a third planet as a function of the orbital period of that planet. Offsets between data from different instruments and inhomogeneous  measurement uncertainties are incorporated into $\chi^2$. In order to assess the significance of peaks in the periodogram we determine an empirical false alarm probability (eFAP) by fitting a log-linear function to the distribution of values in a given periodogram.

\begin{deluxetable*}{lrrrr}
\tablecaption{Model Comparison. \label{tab:comp}}
%\tablewidth{245pt}
\tablehead{
	\colhead{Statistic} & \colhead{0 planets} & \colhead{1 planets} & \colhead{2 planets} & \colhead{{\bf 3 planets (adopted)}}
}
\startdata
$N_{\rm data}$ (number of measurements)  & 252 & 252 & 252 & 252\\
$N_{\rm free}$ (number of free parameters)  & 6 & 9 & 14 & 19\\
RMS (RMS of residuals in m s$^{-1}$)  & 4.71 & 4.22 & 3.52 & 3.16\\
$\chi^{2}$ (assuming no jitter)  & 770.54 & 573.94 & 450.02 & 293.6\\
$\chi^{2}_{\nu}$ (assuming no jitter)  & 3.13 & 2.36 & 1.89 & 1.26\\
$\ln{\mathcal{L}}$ (natural log of the likelihood)  & -736.59 & -701.96 & -662.21 & -621.8\\
BIC (Bayesian information criterion)  & 1484.71 & 1418.45 & 1343.95 & 1268.13\\
\enddata
\end{deluxetable*}

\begin{deluxetable*}{lrl}
\tablecaption{The MCMC posterior values for the three-planet solution. The measured system velocity ($\gamma$) and the derived jitter term ($\sigma_{\rm{jit}}$) are quoted for each of the three instruments. \label{tab:params}}
\tablehead{\colhead{Parameter} & \colhead{Value} & \colhead{Units}}
\startdata
\sidehead{\bf{Orbital Parameters}}
$P_{b}$ & 0.959641 $\pm 1.1e-05$ & days\\
$T\rm{conj}_{b}$ & 2457394.37454 $\pm 0.00044$ & JD\\
$e_{b}$ & $\equiv$ 0.0  & \\
$\omega_{b}$ & $\equiv$ 0.0  & radians\\
$K_{b}$ & 3.58 $^{+0.26}_{-0.25}$ & m s$^{-1}$\\
$P_{c}$ & 29.8454 $\pm 0.0012$ & days\\
$T\rm{conj}_{c}$ & 2457394.9787 $^{+0.0012}_{-0.0011}$ & JD\\
$e_{c}$ & $<$0.267 & \\
$\omega_{c}$ & -3.2 $^{+2.0}_{-1.9}$ & radians\\
$K_{c}$ & 2.24 $\pm 0.28$ & m s$^{-1}$\\
$P_{d}$ & 8.492 $^{+0.023}_{-0.024}$ & days\\
$T\rm{conj}_{d}$ & 2457806.1 $\pm 0.5$ & JD\\
$e_{d}$ & $<$0.36 & \\
$\omega_{d}$ & -3.2 $\pm 1.4$ & radians\\
$K_{d}$ & 2.39 $\pm 0.24$ & m s$^{-1}$\\
\hline
\sidehead{\bf{Modified MCMC Step Parameters}}
$\sqrt{e}\cos{\omega}_{b}$ & $\equiv$ 0.0  & \\
$\sqrt{e}\sin{\omega}_{b}$ & $\equiv$ 0.0  & \\
$\sqrt{e}\cos{\omega}_{c}$ & 0.001 $\pm 0.15$ & \\
$\sqrt{e}\sin{\omega}_{c}$ & 0.01 $\pm 0.24$ & \\
$\sqrt{e}\cos{\omega}_{d}$ & -0.14 $^{+0.23}_{-0.19}$ & \\
$\sqrt{e}\sin{\omega}_{d}$ & 0.002 $\pm 0.23$ & \\
\hline
\sidehead{\bf{Other Parameters}}
$\gamma_{\rm HIRES}$ & -0.9 $^{+0.46}_{-0.47}$ & m s$-1$\\
$\gamma_{\rm APF}$ & -0.51 $^{+0.36}_{-0.37}$ & m s$-1$\\
$\gamma_{\rm HARPSN}$ & 19528.8 $\pm 0.23$ & m s$-1$\\
$\dot{\gamma}$ & $\equiv$ 0.0  & m s$^{-1}$ day$^{-1}$\\
$\ddot{\gamma}$ & $\equiv$ 0.0  & m s$^{-1}$ day$^{-2}$\\
$\sigma_{\rm HIRES}$ & 3.42 $^{+0.4}_{-0.35}$ & $\rm m\ s^{-1}$\\
$\sigma_{\rm APF}$ & 3.45 $^{+0.3}_{-0.27}$ & $\rm m\ s^{-1}$\\
$\sigma_{\rm HARPSN}$ & 1.4 $^{+0.22}_{-0.19}$ & $\rm m\ s^{-1}$\\
\enddata
\tablenotetext{}{The reference epoch for $\gamma$,$\dot{\gamma}$,$\ddot{\gamma}$ is 2457652.6.}
\tablenotetext{}{507,500 links were saved.}
\end{deluxetable*}

\begin{figure*}[t]
\centering
\includegraphics[width=0.76\textwidth]{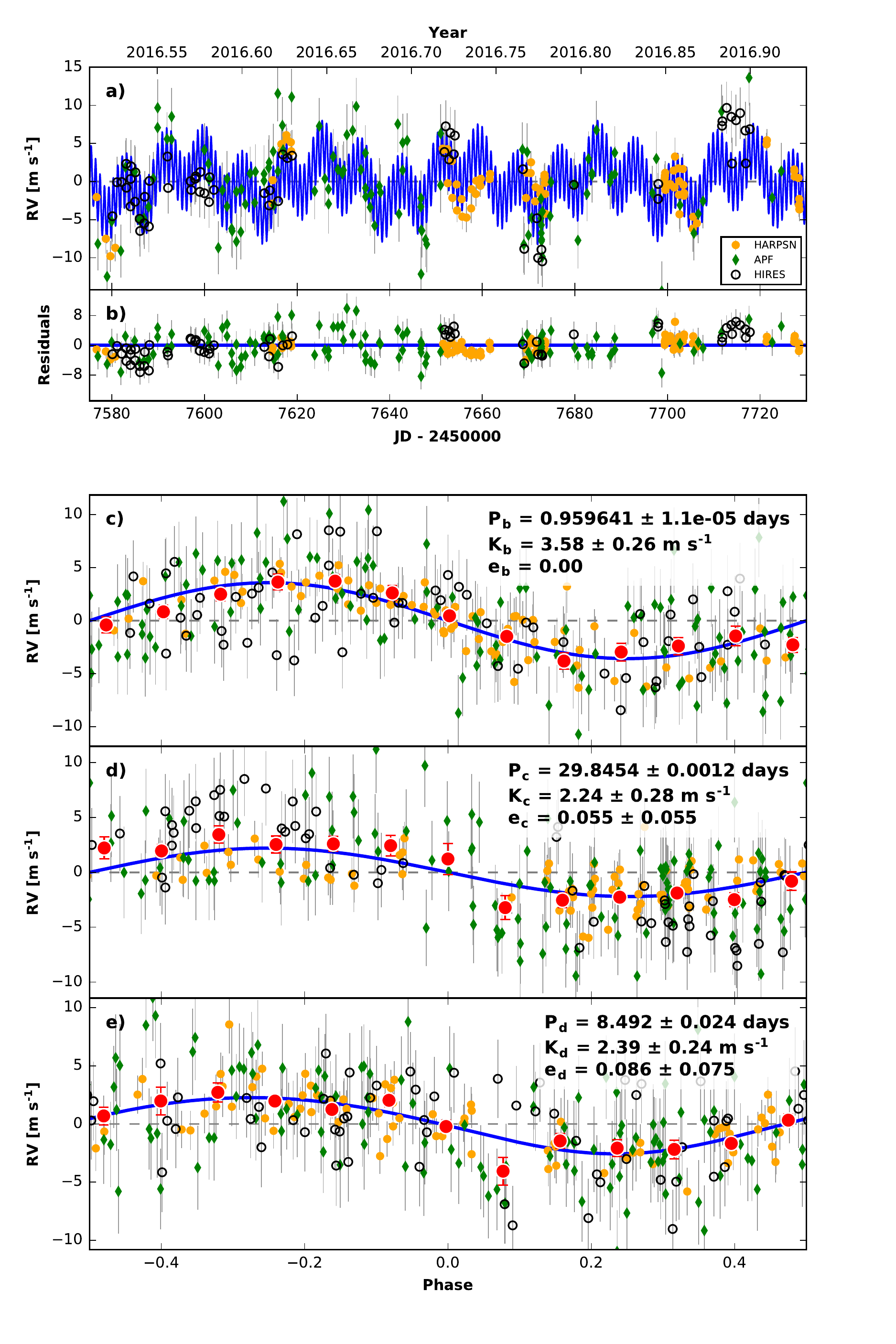}
\caption{{\bf a)} The best-fit three-planet Keplerian orbital model for HD~3167. In each panel, the yellow circles are the HARPS-N data, the green diamonds are the APF data, the open black circles are the HIRES data, and the red circles are the binned data. The maximum likelihood model is plotted; the orbital parameters listed in Table \ref{tab:params} are the median values of the posterior distributions. The thin blue line is the best fit 3-planet model. The uncertainties plotted include the RV jitter term(s) listed in Table \ref{tab:params} added in quadrature with the measurement uncertainties for all RVs. 
{\bf b)} Residuals to the best fit 3-planet model.
{\bf c)} RVs phase-folded to the ephemeris of planet b. The Keplerian orbital models for the other planets have been subtracted.
The small point colors and symbols are the same as in panel {\bf a}.
The red circles are the same velocities binned in units of 0.08 of the orbital phase.
The phase-folded model for planet b is shown as the blue line.
Panels {\bf d)} and {\bf e)} are the same as panel {\bf c)} but for planets c and d respectively. \label{fig:rvmultipanel}}
\end{figure*}

\begin{figure*}[t]
\centering
\includegraphics[width=0.65\textwidth]{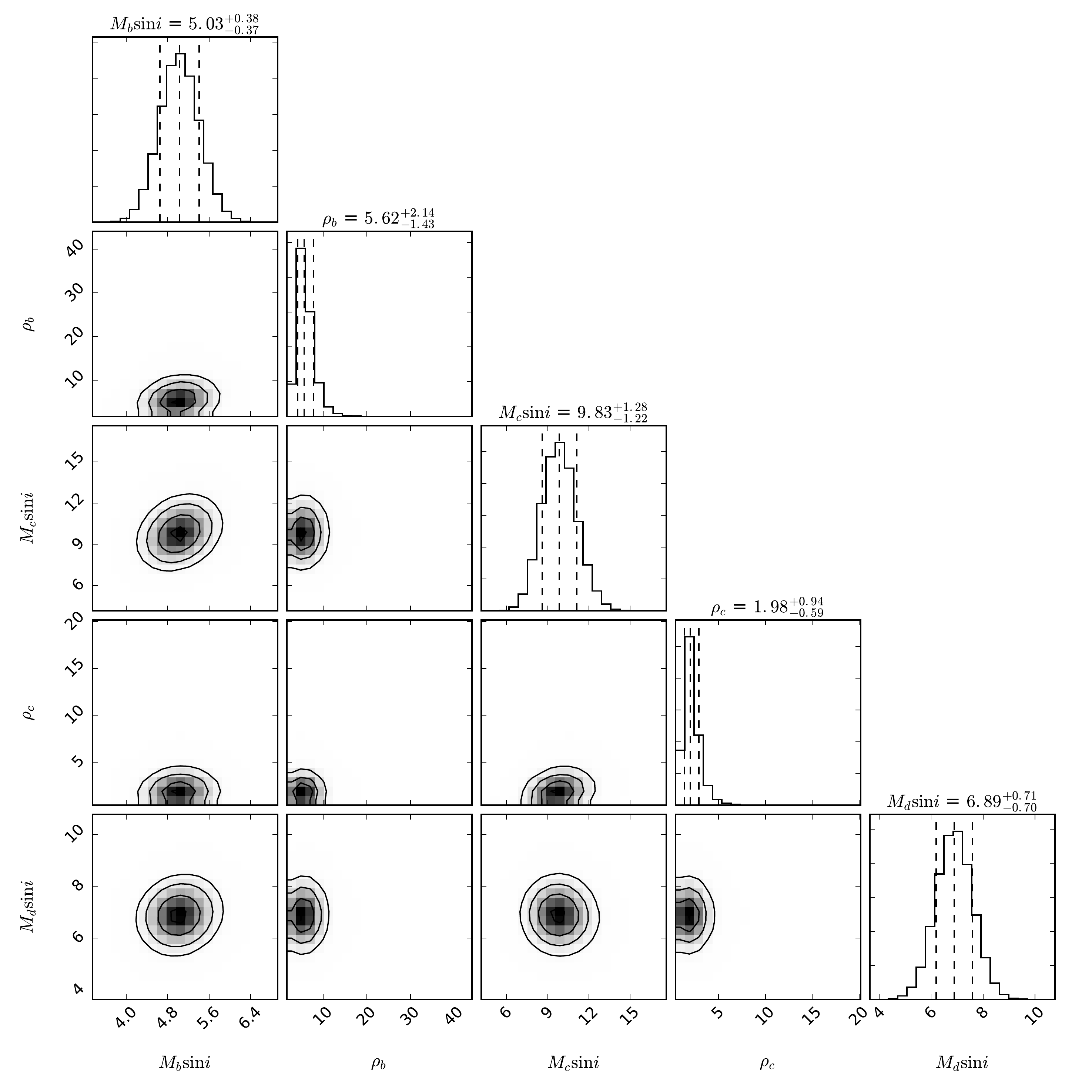}
\caption{The correlations between the derived parameters in the three-planet Keplerian orbital model. The marginally incomplete phase combination coverage between planets b and c, shown in Figure \ref{fig:phases}, manifests as a slight degeneracy between the masses of the two planets. The more incomplete the coverage, the higher the resulting degeneracy. \label{fig:corner}}
\end{figure*}

\begin{figure}[t]
%\epsscale{1.25}
%\plotone{3167_multi-peri.pdf}
\includegraphics[width=\columnwidth]{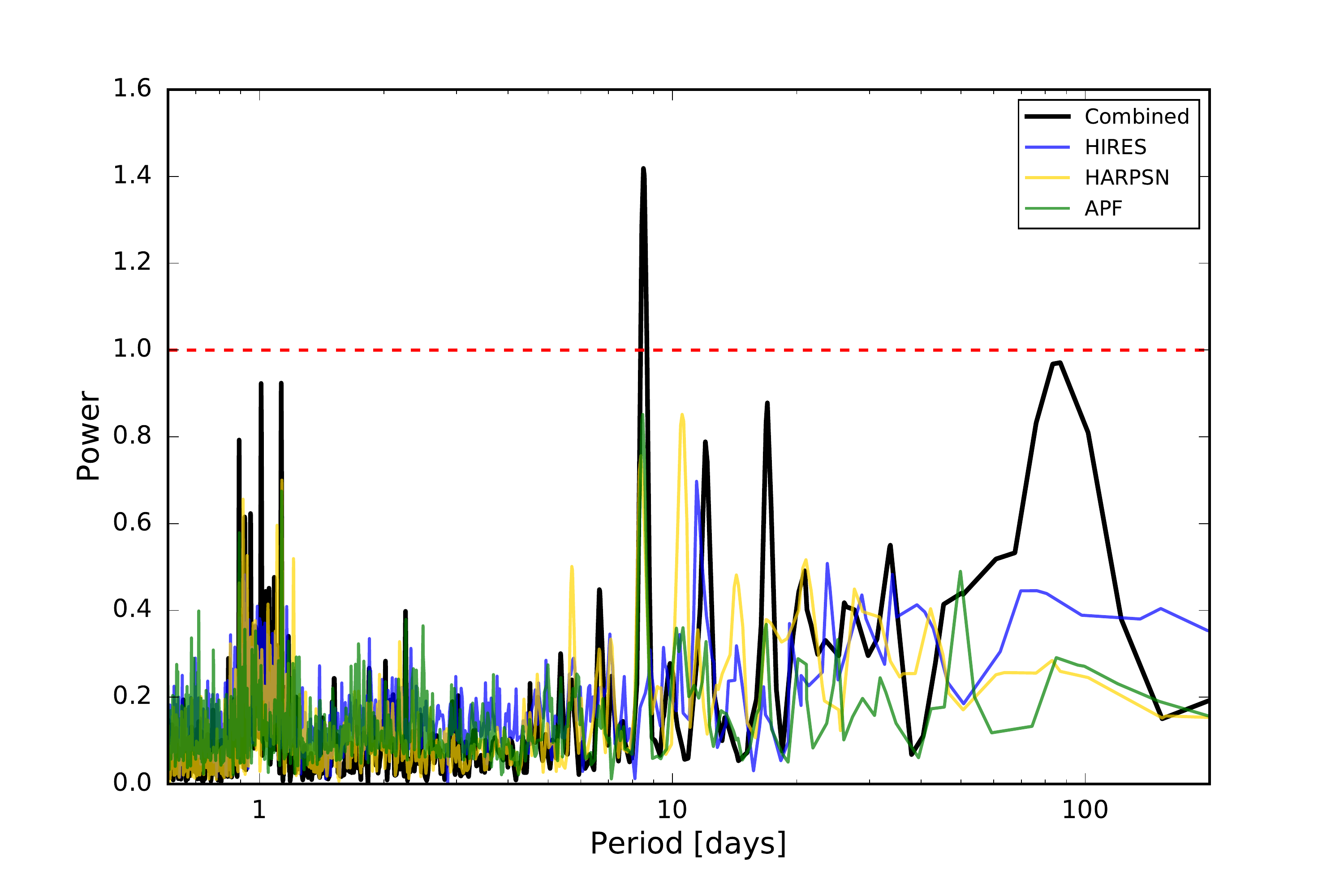}
\includegraphics[width=\columnwidth]{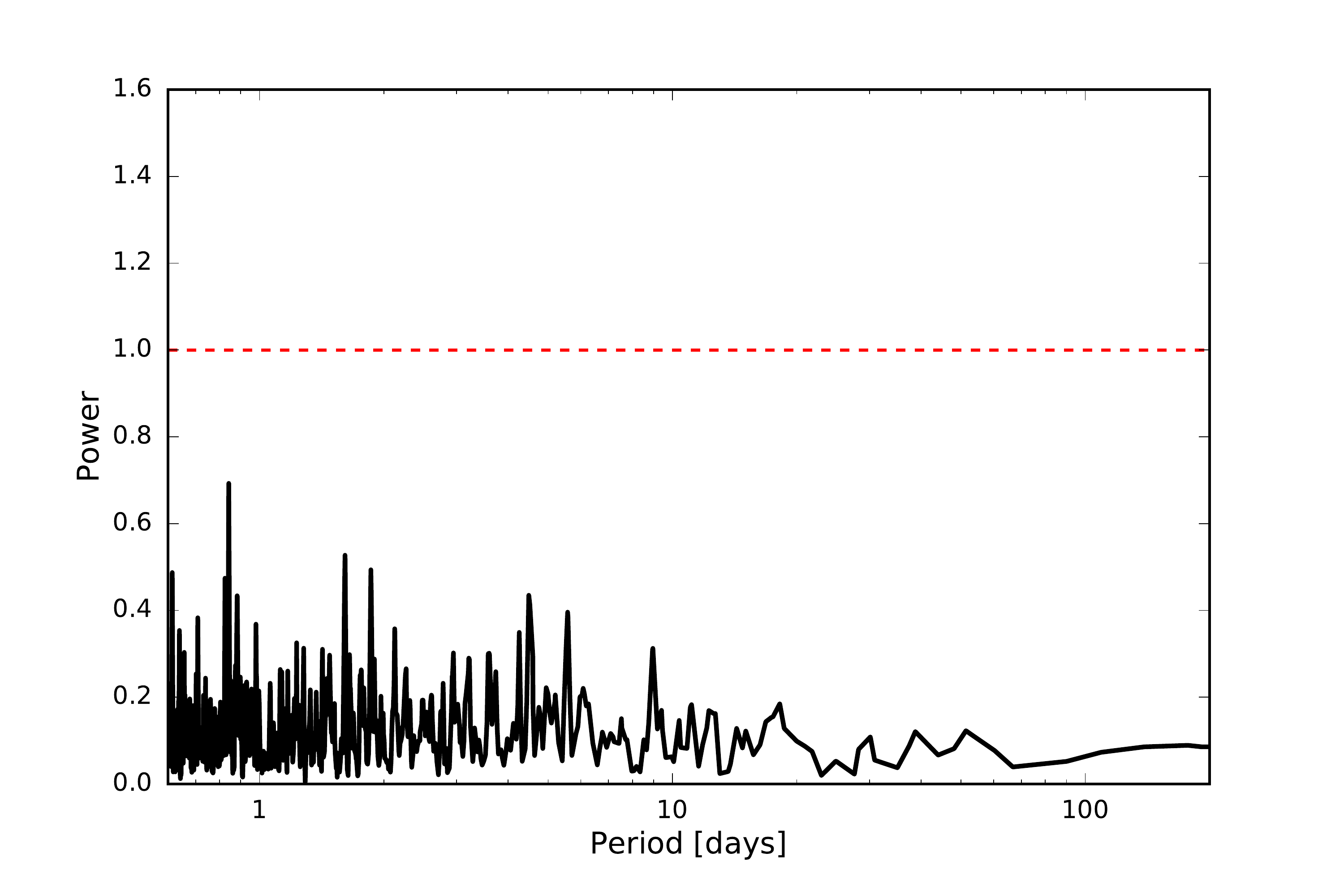}
\caption{\emph{Top panel}: 2DKLS periodogram of the combined RV data showing the improvement to $\chi^2$ for a three planet fit relative to that of a two planet fit (thick black line). We find a significant peak with eFAP$\approx$0.3\% at an orbital period of 8.5 days. Periodograms of the HIRES, HARPSN, and APF data independently are shown in blue, gold, and green respectively. All periodograms have been normalized such that power$=$1.0 is equivalent to eFAP$=$1\% (also indicated by the red dashed line). \emph{Bottom panel}: 2DKLS periodogram of the simulated radial velocity curve containing the two transiting planets and preserving the observing window function, after removal of the two known signals.}
\label{fig:multi-peri}
\end{figure}

We find a significant peak with eFAP$\sim$0.3\% and a period of $\sim$8.5 days in the 2DKLS periodogram of the combined RV dataset  when searching for a third Keplerian signal. When we add this additional Keplerian into the MCMC fits described in Section \ref{sec:rvanalysis} we see an improvement in the Bayesian Information Criterion \citep[BIC,][]{Liddle07} of 76, which indicates that the three-planet model is highly favored over the two-planet model.

We also calculate the 2DKLS periodogram for each instrument independently. In the 2DKLS periodogram for the APF data, we find that the highest periodogram value similarly falls at a period of $\sim$8.5 days, with an eFAP$\sim$20\%. We find that the highest peak in the 2DKLS periodogram of the HARPS-N data falls at a period of $\sim$11 days, which is near an alias of 8.5 days caused by the sampling being concentrated around lunar cycles ($1/8.5 \rm{~days} - 1/29.5 \rm{~days} = 1/11.9 \rm{~days}$). The second highest peak in the 2DKLS periodogram of the HARPS-N data falls at a period of 8.4 days. The HIRES data also shows an insignificant peak with a period of $\sim$11 days. The APF data, which has much more uniform sampling due to the semi-dedicated nature of the telescope, is critical to break the monthly alias and reveal the true period of the third planet.

In order to examine whether the 8.5-day signal could be caused by a window function effect, in the fashion of $\alpha$ Cen Bb \citep{Rajpaul2016}, we perform the following test: using the real observing times, we generate a simulated radial velocity curve from the properties of the two transiting planets. For each point, we generate an uncertainty drawn from a normal distribution of the quadrature sum of the observation error and the instrument jitter for the instrument that obtained that observation. We run the simulated radial velocity curve through 2DKLS, and after removing the two known signals, we see no significant remaining power in the 7--10-day range, implying that the observed 8.5-day signal in the real data is not caused by a window function effect of the observations. We show the results of this final search in the lower panel of Figure \ref{fig:multi-peri}. 

We also examined whether the 8.5-day signal could be caused by stellar activity, since the period is potentially near an integer alias of the stellar rotation period. The best-fit jitter value for Keck/HIRES is surprisingly large in comparison to that from the HARPS-N dataset. Long term Keck/HIRES monitoring of stars with similar spectral types and activity levels show jitter as low as 1.8 m/s. Inspection of the residuals in Figure \ref{fig:rvmultipanel} show a systematic structure that appears to be present in only the Keck/HIRES dataset. These correlated residuals are the source of the inflated jitter. We collected iodine-free template observations for this star on three different occasions and recalculated the velocity time series using each of the different templates. The results were comparable in each case and the structure in the residuals did not change significantly. We also searched for correlations of the velocity residuals with environmental and pipeline parameters. The Keck/HIRES velocity residuals are weakly correlated with both barycentric correction and S value. We tried subtracting a linear trend from radial velocity against barycentric correction and/or S value by adding a term into the likelihood in the MCMC fit, but found only very modest improvement to the final jitter value and no significant difference to the final results. Since the structure in the residuals appears to be quasi-periodic and weakly correlated with S value, we suspect that the source of the large jitter is likely caused by rotational modulation of starspots. The iodine technique used to extract the velocities from the Keck/HIRES and APF spectra could be more sensitive to the line-shape distortions produced by these starspots compared to the cross-correlation technique used to extract the velocities from the HARPS-N spectra. As shown in Figure \ref{fig:multi-peri}, the signal of HD~3167 d is present in the HARPS-N and APF data, which do not show systematic structure in their residuals, so we are confident that the signal of planet d is not caused by stellar activity. We investigated this further by examining the stacked periodogram of the radial velocities \citep{Mortier2017} and noting that the strength of the 8.5-day signal peak in the periodogram increases with the addition of more data, as distinct from the behaviour of a peak caused by quasi-periodic stellar activity.

\subsection{Composition}
\label{sec:composition}

\begin{figure}[t]
\centering
%\figurenum{9}
\includegraphics[width=\columnwidth]{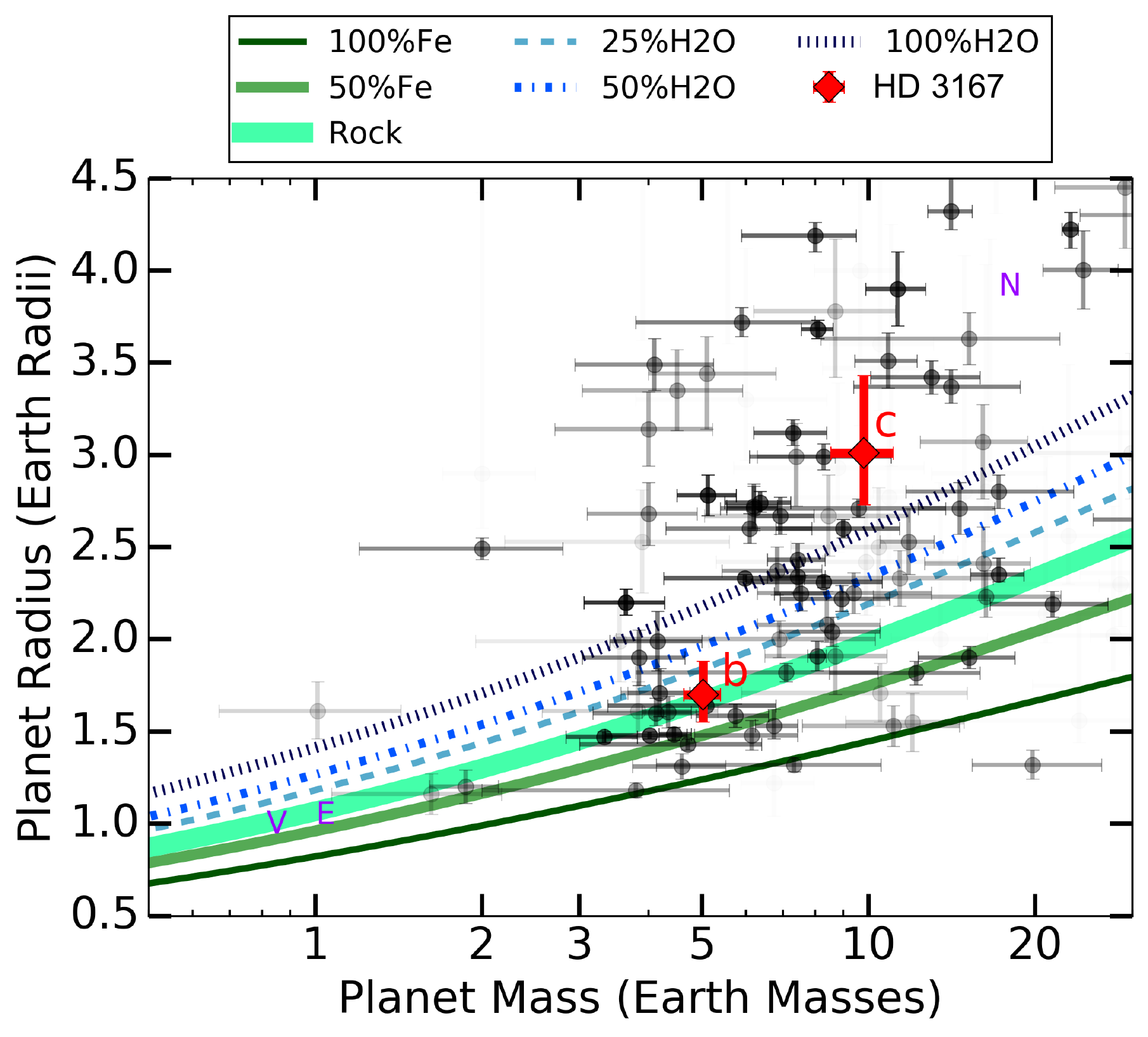}
%\plotone{hd3167_mass_radius_50percentmass-1.png}
%\plotone{hd3167_mass_radius_20percentmass.png
\caption{Masses and radii for planets with masses measured to better than 50\% uncertainty. The shading of the points and error bars corresponds to their uncertainty---darker points are more precisely constrained. The red points are the newly added HD~3167 b and c values from this paper. N, V, and E mark the solar system planets. The curves show the mass-radius correlation for compositions ranging from 100\% iron to 100\% water from \citet{Zeng2016}. Planet b is likely predominately rocky, and planet c is volatile-rich.\label{fig:mrdiagram}}
\end{figure}

The measured mass and radius of HD~3167~b (\pmassberr~M$_{\oplus}$, \pradb~R$_{\oplus}$) indicate a bulk density of \pdensberr\ g~cm$^{-3}$; consistent with a predominantly rocky composition, but potentially having a thin envelope of H/He or other low-density volatiles. Figure \ref{fig:mrdiagram} shows HD~3167~b in comparison with other small exoplanets with masses measured to better than 50\% precision; the lines show the composition models of \citet{Zeng2016}. We randomly draw 100,000 planet masses and radii from our posterior distributions, and compare them to the mass-radius relation of \citet{Fortney2007} for pure rock, finding results that are consistent with the models of \citet{Zeng2016}. Assuming that the planet is a mixture of rock and iron, we compute the iron mass fraction from each random draw using Equation 8 of \citet{Fortney2007}. We conclude that the iron mass fraction is smaller than 15\% at 68\% confidence and smaller than Earth's iron mass fraction (33\%) at 85\% confidence, under the assumption that the planet is a mixture of rock and iron, with no volatiles. The radius, \pradb~R$_{\oplus}$, brackets the putative transition radius from likely rocky to likely volatile rich at 1.6~R$_{\oplus}$ proposed by \citet{Rogers2015}. Planetary envelopes in such close proximity to the host star are predicted to be stripped away, either through photo-evaporation \citep[e.g.][]{Owen2012, Lopez2014, Chen2016, Lopez2016} or Roche lobe overflow \citep[e.g.][]{Valsecchi2014}. Our constraints are consistent with the notion that ultra-short-period (USP) planets are predominantly rocky. %However, it is unclear whether or not HD~3167~b has a thin, volatile-rich envelope.

%We note that the mass and radius of HD 3167~b are consistent with another USP planet CoRoT-7~b, which has a mass of 4.73$\pm$ 0.95~M$_\oplus$ \citep{Haywood2014} and radius of 1.585$\pm$0.064~R$_\oplus$ \citep{Barros2014}. CoRoT-7 also hosts at least one additional non-transiting planet, CoRoT-7c ($P$=3.7~days) with M$_{p}\sin{i}$= 13.56$\pm$1.08~M$_\oplus$ \citep{Haywood2014} suggesting these two systems could have similar formation histories. It is unknown whether the orbits of CoRoT-7b and CoRoT-7c are mutually inclined like the inner planets of HD 3167. The absence of CoRoT-7c transits can be explained by a non-coplanar orbit or its larger orbital distance. 
 
HD~3167~c has a mass and radius of \pmasscerr~M$_{\oplus}$ and \pradc~R$_{\oplus}$ respectively, also shown in Figure \ref{fig:mrdiagram}. The resulting bulk density of HD~3167~c is \pdenscerr\ g cm$^{-3}$. The mass and radius can be explained by a wide range of compositions, all of which include low-density volatiles such as water and H/He \citep{Adams2008, Rogers2010, Valencia2013}. The planet evolution models of \citet{Lopez2016} are consistent with an Earth-composition core surrounded by a H/He envelope comprising $\sim$2\% of the total planet mass. Alternatively, the planet might be mostly water. With a K-band magnitude of 7, HD~3167 is amenable to transmission spectroscopy observations to detect the atmospheric constituents of planet c, discussed in Section \ref{sec:jwst}, which will help to break compositional degeneracies. HD 3167~c receives an incident flux $\approx$16 times that of Earth, and is much less susceptible to atmospheric photo-evaporation than planet b.  Planet b could be a remnant core of a planet similar to planet c. 

\section{Prospects for atmospheric study}
\label{sec:jwst}

The brightness of the host makes the planets HD~3167~b and c excellent candidates for detailed atmospheric characterization. The low bulk
density of planet c, in particular, suggests that the planet is surrounded by a thick gas envelope, as discussed in Section \ref{sec:composition}. %Atmospheric modeling shows that the amplitude of water absorption features in the transmission spectrum of the atmosphere of planet c could reach up to 100~ppm if the atmosphere is hydrogen-rich and cloud free (see Figure \ref{fig:JWST}). Such a signature would readily be detectable by stacking 2-3 transit observations with \emph{HST}/WFC3 due to the brightness of the star. 
If HD~3167~c has a large extended exosphere, HST/UV observations could detect escaping hydrogen, as for GJ~436b \citep{Ehrenreich2015}. Beyond current instrumentation, \emph{JWST}/NIRISS would simultaneously observe 0.6 to 2.8~$\mu$m and provide robust detections of all main water absorption bands in the near-infrared. Here, we estimate that an NIRISS SOSS spectrum would provide near photon-noise-limited observations, with approximately 15 ppm uncertainty when binned to $R=100$ at $\lambda = 1.2-1.8\mu$m. Molecular detections for high-metallicity atmospheres or hydrogen-rich atmospheres with high-altitude clouds above 1~mbar will, however, be substantially be more challenging due to the lower signal-to-noise afforded by the relatively large stellar radius  \citep{Benneke2013}. We estimate that a robust distinction between an atmosphere with a high mean molecular weight and a cloudy hydrogen-dominated atmosphere with solar water abundance would require multiple \emph{JWST} visits.

HD~3167~b, on the other hand, is likely to have been stripped of a substantial volatile component due to its proximity to the host star. However, the higher equilibrium temperature of planet b makes it the better target for secondary eclipse observations of its thermal emission, despite its smaller radius and shorter transit time. Given that its short P$<$1 d orbit is unlikely to be significantly eccentric, we assume that its secondary eclipse duration equals its transit duration and we can expect the eclipse to occur at mid-time between transits. Assuming a planetary equilibrium temperatures of 1700~K for planet b and approximating the planet as blackbody we would expect a thermal emission signal ($F_{p}/F_{s}$) of $\gtrsim60$~ppm longward of 5~$\mu m$. We estimate that the thermal emission of this planet could be detected at S/N $\simeq$ 8 in a single secondary eclipse observation at wavelengths 4\textendash 7 $\mu$m with a $R=4$ filter if only photon noise is considered. Introducing only 20~ppm of systematic noise would reduce this to S/N $\simeq$ 3, so this will likely be a difficult observation. The $\lambda<5\mu$m \emph{JWST} NIRCam detectors will likely have lower residual systematic noise than the $\lambda>5\mu$m MIRI ones \citep{Beichman2014}, so an observation with the NIRCam F444W filter may be the best way to detect this signal. This and all other calculations assume equal time spent observing the star HD 3167 alone outside of transit or secondary eclipse. 

\begin{figure}[t]
%\figurenum{12}
\includegraphics[width=\columnwidth]{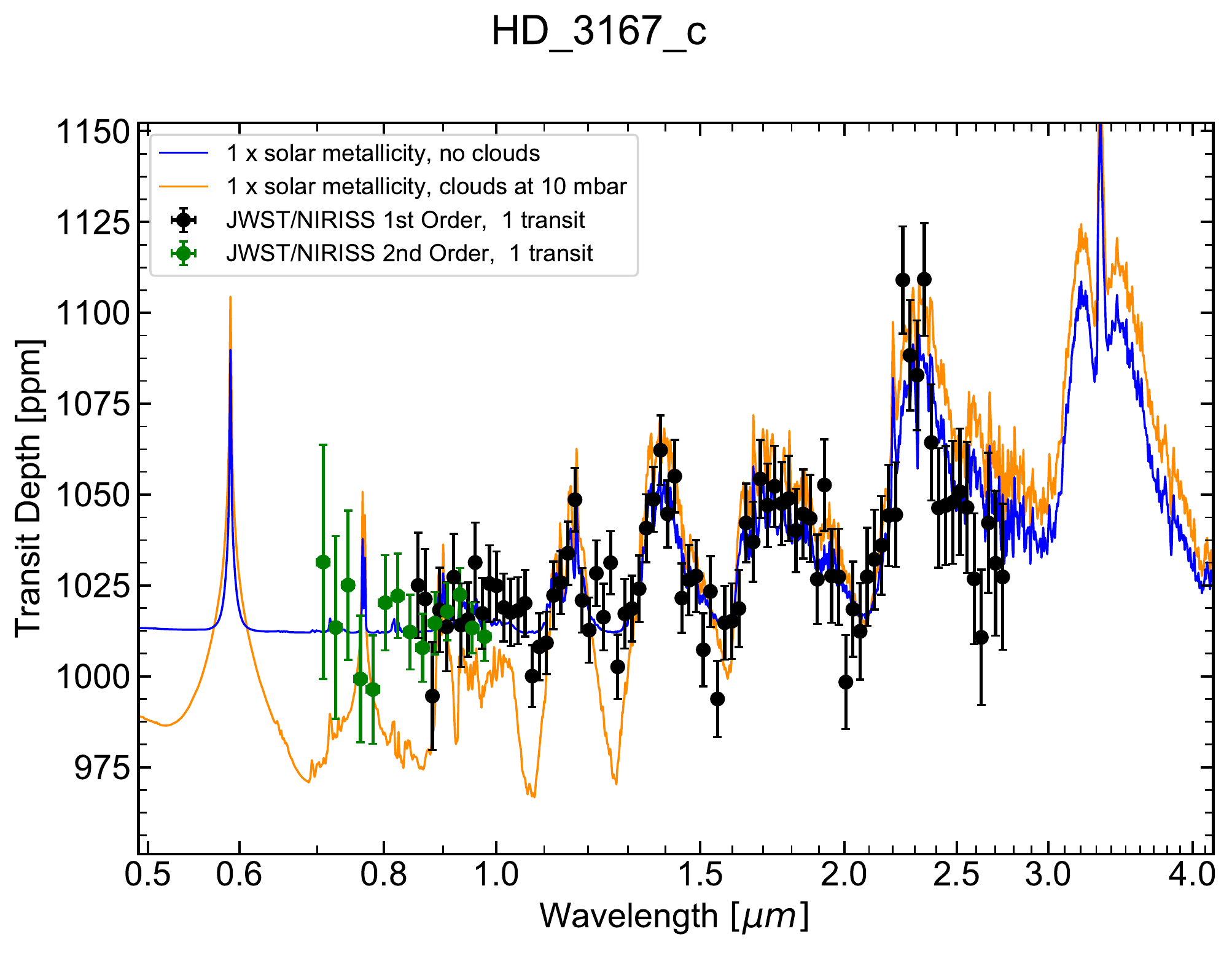}
\caption{Model transmission spectra and simulated observations of the mini-Neptune HD~3167~c, binned to $R=70$ in the first order, and $R=40$ in the second order. Assuming a single transit observation by \emph{JWST}, water absorption is detectable at high significance in both cloud-free and cloudy scenarios. Models were generated as described in \citep{Benneke2012,Benneke2015}. The observational uncertainties are 120\% of the photon-noise limit accounting for the exact throughput, duty-cycle, and dispersion of the instruments.\label{fig:JWST}}
\end{figure}

\section{Dynamics}
\label{sec:dynamics}
 
In this section, we consider the dynamical behavior of the three planet system with an eye towards placing additional constraints on its orbital architecture. The architecture is notable due to the misalignment of the middle planet compared to the coplanar inner and outer planets. We begin by noting that the optimal fit to the combined data set yields a period ratio of planets c and d that is very close to $7/2$. In light of this near-commensurability, it is worthwhile to inspect the possibility that the \textit{c-d} planetary pair is currently locked in a $7/2$ mean-motion resonance (MMR). We note that although the $7/2$ commensurability arises at 5th order in the perturbation series \citep[hereafter MD99]{Murray1999}, at least one example of an extrasolar planetary system, Kepler-36 \citep{Deck2012}, is known to currently reside in a 5th order (29:34) MMR.

\subsection{Mean Motion Commensurability}

Unlike the case of Kepler-36, with an orbit tightly constrained by transit timing variations, the radial velocity orbital fit of HD~3167 is not sufficiently precise to deduce the behavior of resonant harmonics directly. Thus, we approach this question from an alternative viewpoint---namely, we employ numerical experiments to examine whether the conditions required to establish such a resonant lock could have occurred in the system's evolutionary history. It is well known that mean-motion resonances arise from smooth convergent migration (in this case, likely due to interactions with the protoplanetary nebula), and the probability of capture depends both on the planetary eccentricities at the time of the resonant encounter, as well as the migration rate \citep{Henrard1982,Borderies1984}. Application of adiabatic theory \citep{Neishtadt1975} shows that resonance capture probability diminishes with increasing eccentricity and/or increasing migration rate \citep{Batygin2015}. Accordingly, in our simulations, we circumvent the former issue by assuming that the planets approach one another on initially circular orbits, and only retain the migration rate as an adjustable parameter.

To facilitate orbital convergence and damping, we have augmented a standard gravitational \textit{N}-body code with fictitious accelerations of the form \citep{Papaloizou2000}: 
\begin{equation}
\frac{d\vec{v}}{dt} = -\frac{\vec{v}}{\tau_{\rm{mig}}} - \frac{2\,\vec{r}}{\tau_{\rm{dmp}}} \frac{\left(\vec{v}\cdot{\vec{r}}\right)}{\left(\vec{r} \cdot \vec{r} \right)},
\label{PapLar}
\end{equation}
where $\tau_{\rm{mig}}$ and $\tau_{\rm{dmp}}$ are the migration and damping timescales, respectively. For definitiveness, migration torque was only applied to the outer planet, while damping torques were exerted upon both planets. Additionally, the gravitational potential of the central star was modified to account for the leading-order effects of general relativity \citep{Nobili1986}. The simulations employed the Bulirsch-Stoer algorithm \citep{Press1992}, and initialized the orbits in the plane, with random mean anomalies, $\sim5\%$ outside of the exact $7/2$ resonance. 

We have carried out a sequence of numerical experiments with $\tau_{\rm{mig}}$ ranging from the nominal type-I migration timescale of $\sim 5000\,$years \citep{Tanaka2002} to $\tau_{\rm{mig}}=3\,$Myr \citep[i.e. a typical protoplanetary disk lifetime;][]{Armitage2010}, and with $\tau_{\rm{dmp}}=\infty$ as well as $\tau_{\rm{dmp}}=\tau_{\rm{mig}}/100$ \citep{Lee2002}. We tested each parameter combination with ten cloned simulations, and did not observe capture into a $7/2$ MMR a single time. As a consequence, we conclude that it is unlikely that the planets are presently affected by the nearby $7/2$ resonance, and the orbital proximity to this commensurability is coincidental. 

\subsection{Lagrange-Laplace Theory}

\begin{figure*}[ht!]
%\figurenum{10}
\includegraphics[width=\textwidth]{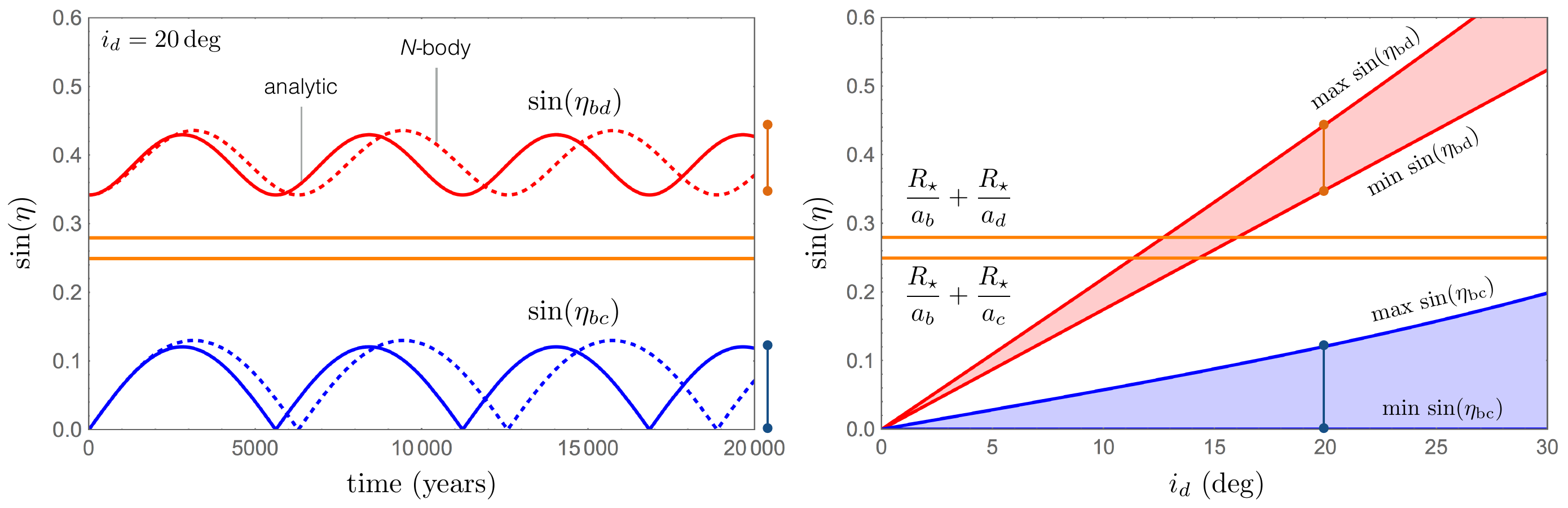}
\caption{Evolution of mutual inclinations within the HD 3167 system. The left panel depicts the sine of mutual inclinations of planets b and c (blue) as well as that corresponding to planets b and d (red), adopting a present-day inclination of planet d of $i_d = 20$ deg. The solid and dashed curves correspond to solutions computed analytically (solid) and using a direct N-body approach (dashed). The two orange lines show critical misalignments, given by equation \ref{eq:sinlimit}. The right panel depicts the range of mutual misalignments attained by the planet pairs (color-coded in the same way) as a function of planet d's present-day inclination. While an inclination in excess of $i_d >1.3$ deg will allow planet d to not transit given a favorable alignment of the nodes, an inclination of $i_d >15$ deg is required to reproduce the architecture of the system without invoking a specific nodal configuration.\label{LLfig}}
\end{figure*}

With the possibility of resonant interactions disfavored, we proceed with a purely secular (i.e. orbit-averaged) treatment of the 
dynamics. A specific question we now seek to address concerns the mutual inclinations within the system. In other words, what extent of misalignment among the angular momentum vectors of the planetary orbits is required for planet d to elude transit, 
while allowing planets b and c to transit simultaneously? Although an exact answer to this question can in principle be attained from numerical integrations, such calculations require a more precise knowledge of the input parameters (e.g. eccentricities, longitudes of periastron, etc) than what is presently available. Consequently, here we settle for an approximate answer, which we deduce analytically from secular perturbation theory. %Correspondingly, \textit{N}-body simulations will only be employed as a check on our calculations.

A conventional approach to modeling the long-term behavior of planetary systems that reside outside of mean-motion commensurabilities, is to replace the planetary orbits with massive wires and compute the resulting exchange of angular momentum (MD99). We note that formally, this is equivalent to averaging the governing Hamiltonian over the mean longitudes \citep{Morbidelli2002}. In the limit of low eccentricities and mutual inclinations (specifically, to second order in either quality), the inclination and eccentricity dynamics become decoupled, meaning that the uncertainties of the RV fit do not strongly affect the following calculations. 

Within the context of this so-called Lagrange-Laplace secular theory \citep[see][for a complete discussion]{Brouwer1961}, the equations of motion for the complex inclination vector $z=i\,\exp(\imath\,\Omega)$, where $i$ is the inclination and $\Omega$ is the ascending node, simplify to a linear eigenvalue problem:
\begin{equation}
\frac{dz_{j}}{dt} = \imath \sum_{k=1}^N B_{jk} z_{k},
\label{eigenvalue}
\end{equation}
where the indexes run over the planets, and $N=3$. The interaction coefficients $B_{jk}$ depend exclusively on the planetary masses as well as the semi-major axis ratios, and comprise a matrix \textbf{B} that fully encapsulates the dynamics:
\begin{align}
B_{jj} &=-\frac{n_{j}}{4}\sum_{k=1,k\neq j}^N \frac{m_k}{M_{\star}} \alpha_{jk} \bar{\alpha}_{jk} b^{(1)}_{3/2}(\alpha_{jk})  \nonumber \\
B_{jk} &=\frac{n_{j}}{4}\frac{m_k}{M_{\star}} \alpha_{jk} \bar{\alpha}_{jk} b^{(1)}_{3/2}(\alpha_{jk}).
\end{align}
In the above expression, $n=\sqrt{GM_{\star}/a^3}$ is the mean orbital frequency, $\alpha < 1$ is the semi-major axis ratio, $b^{(1)}_{3/2}(\alpha_{jk})$ is a Laplace coefficient of the first kind, and $\bar{\alpha} = \alpha$ if $a_j<a_k$; $\bar{\alpha} = 1$ if $a_k<a_j$. With these specifications of the problem, the solution to equation (\ref{eigenvalue}) can be be expressed as a super-position of \textit{N} linear modes:
\begin{equation}
z_{j} = \sum_{k=1}^N \beta_{jk} \exp(\imath f_k t + \delta_k),
\end{equation}
where $f_k$ and $\beta_{jk}$ denote the eigenvalues and eigenvectors of \textbf{B}, respectively. The scaled amplitudes of the eigenvectors and the phases $\delta_k$ are determined entirely by the specific choice of initial conditions.

For definitiveness, here we initialize the transiting planets (b and c) in the plane ($i_b=i_c=0$; $\Omega_b,\Omega_c$ undefined), and choose our reference direction to coincide with the present-day ascending node of the inclined planet d ($\Omega_d=0$). Although adopting this initial condition does not lead to a general analysis of the systems?s possible dynamical evolution, this simplification is justified given the current observational constraints. Consequently, the only free parameter that enters our calculations is planet d's inclination. Moreover, owing to the analytic nature of our solution, the computational cost associated with any one realization of the dynamics is negligible. 

To obtain an absolute lower-bound on planet d's present-day inclination, we note that given a favorable configuration of the line of nodes relative to the line of sight, any inclination greater than $i_d > \rm{arctan} (R_{\star} / a_d)=1.3\deg$ will allow planet d to elude transit for some fraction of the time, potentially during the 80-day duration of the \emph{K2} observations. The greater the mutual inclination, the larger the fraction of time that planet d does not transit, rising from $\sim$7\% for an inclination of 3 degrees, to $\sim$80\% for inclinations of 10 degrees. The nodal configuration assumption therefore becomes progressively less stringent as the adopted value of $i_d$ increases, and it is of interest to estimate the critical $i_d$ beyond which this limitation can be alleviated altogether\footnote{Strictly speaking, even for orthogonal orbits, there exists a particular viewing geometry where both planets transit. Practically, however, such configurations are expected to comprise a very small fraction of the observational dataset.}. Moreover, such a calculation can further inform a maximal $i_d$, beyond which none of the planets co-transit.  

Following \citet{Spalding2016}, we define a mutual inclination 
\begin{equation}
\eta_{jk} = \sqrt{z_j z_j^* + z_k z_k^*-( z_j z_k^* + z_k z_j^*) },
\end{equation}
and adopt the following criterion for a pair of planets to co-transit:
\begin{equation}
\sin(\eta_{jk}) < \frac{R_{\star}}{a_j} + \frac{R_{\star}}{a_k}.
\label{eq:sinlimit}
\end{equation}
Generically, as the orbits exchange angular momentum, their mutual inclinations, $\eta_{jk}$, will experience oscillatory motion. An example of this behavior, taking $i_d=20\deg$ as an initial condition, is shown in the left panel of Figure \ref{LLfig}. For reference, the solid lines denote the analytic solutions obtained by matrix inversion, while the dotted lines show the numerical solution computed with the \textit{N}-body code described above. Although a small discrepancy exists in the oscillation frequencies computed analytically and numerically, the amplitudes of oscillation (which are the more relevant quantities for the question at hand) are well captured by secular perturbation theory.

\begin{figure*}[ht!]
%\figurenum{1}
%\plotone{lightcurves.pdf}
\includegraphics[width=\textwidth]{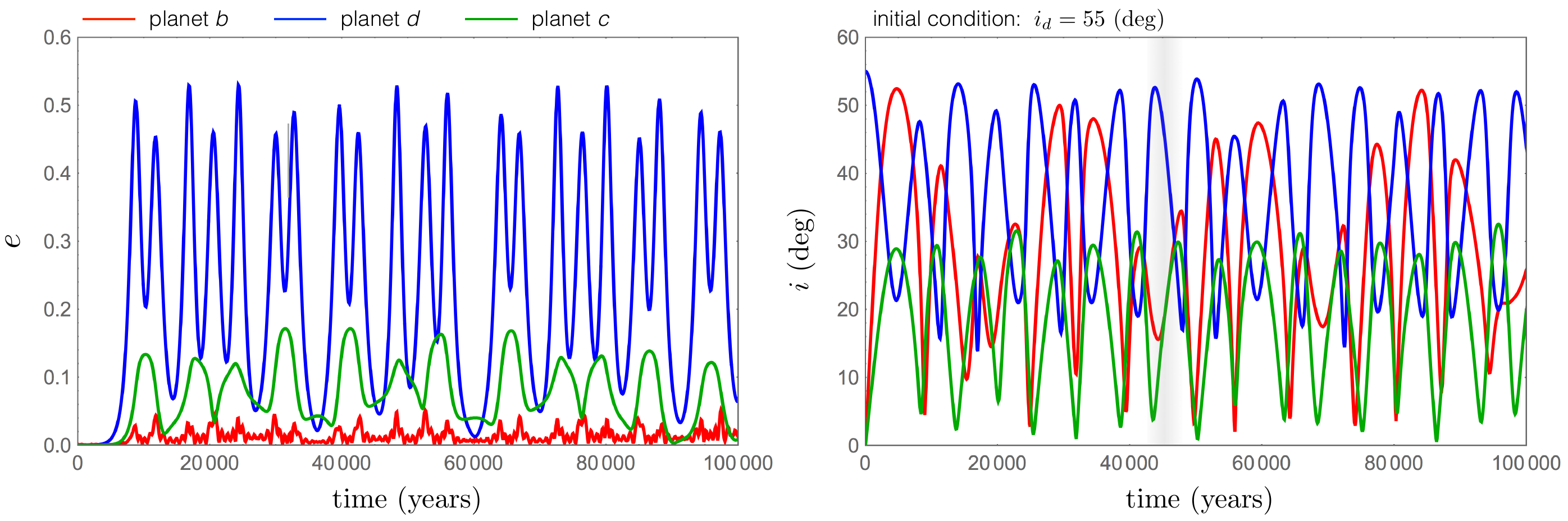}
\caption{Numerically computed evolution of the HD 3167 system in the Kozai-Lidov regime. The left and right panels show eccentricities and inclinations as functions of time, respectively. The red, blue, and green curves correspond to planets b, d and c respectively. The planets are initialized on circular orbits in the plane, with the exception of planet d, which is given an inclination of $i_d=55\deg$. While the system experiences dramatic Kozai-Lidov oscillations, it remains stable indefinitely. Note further, that the approximate recurrence of the initial condition implies that the system periodically returns to a state where planets b and c are essentially coplanar, while planet d possesses a large inclination.\label{KLfig}}
\end{figure*}

In the particular case shown in the left panel of Figure \ref{LLfig}, the orbital architecture of the observed system is correctly reproduced, without assumptions about the current lines of nodes. That is in this case, given almost any nodal configuration, a viewing geometry where planets b and c co-transit, will not permit planets b and d to co-transit also. To estimate the critical inclination of planet d below which all three planets co-transit, we have computed the maximal and minimal extents of mutual inclinations between planets b and c as well as b and d, as a function of $i_d$. These results are shown in the right panel of Figure \ref{LLfig}. Cumulatively, our theoretical calculations suggest that, although planet d can escape transit for inclinations as small as $\sim$1.3 deg, for inclinations above $\sim$15 degrees the allowed range of nodal alignment that would result in planet d transiting becomes so vanishingly small that, in the absence of observed transits, we conclude that the mutual inclination that reproduces the observed orbital misalignment of the HD 3167 system is most likely greater than $\sim$15 degrees.

\subsection{Kozai-Lidov Regime}

While the flavor of secular theory employed above adequately captures the dynamics of the system over the inclination range shown in Figure \ref{LLfig}, the Lagrange-Laplace model is well known to break down at sufficiently high inclinations. Specifically, within the context of the problem at hand, it is reasonable to expect that provided sufficiently large $i_d$, the system will enter the Kozai-Lidov \citep{Lidov1962,Kozai1962} resonance, which can facilitate large-scale oscillations of the eccentricities. A typically quoted inclination, necessary for Kozai-Lidov oscillations to ensue, is $39.2\deg$. Consistently, here we find numerically that when planet d's inclination exceeds $i_d\gtrsim41\deg$, the system enters the Kozai-Lidov regime, and planet d's eccentricity begins to experience oscillations coupled with its argument of pericenter. The small discrepancy in the critical value of the inclination can almost certainly be attributed to the apsidal precession generated by general relativistic effects and the quadrupolar field of the inner planet b \citep{Batygin2011}, as well as the non-negligible mass of planet d itself \citep{Naoz2013}.

Intriguingly, the commencement of Kozai-Lidov oscillations is not synonymous with the onset of dynamical instability. Instead, the system remains stable for at least $100\,$Myr for inclinations up to $i_d\sim60\deg$ (an example of stable evolution with $i_d\sim55\deg$ is shown in Figure \ref{KLfig}). It is only above an inclination of $i_d\sim65\deg$, that eccentricity oscillations become sufficiently extreme, for subsequent orbit crossing to ensue. In this regard, the dynamics of the system entails an observational consequence: if follow-up radial velocity observations sharpen the estimate of planet d's eccentricity to a value that is close to zero, that would imply that planet d's inclination lies below $i_d <40$ deg. Conversely, significant orbital eccentricity in the system would point towards $i_d\simeq41-66\deg$ as the more likely range of orbital misalignment. Constraining the inclination to 15--60 degrees, under the relaxed assumption that requires no special configuration of the lines of nodes, implies a true mass of 7.1--13.8 M$_{\oplus}$ for HD~3167~d.

\subsection{Some Speculation}

The dynamical analysis presented herein shows that the observed orbital architecture of the HD 3167 system can be naturally explained if the orbital inclination of planet d exceeds $\sim15\deg$, without invoking the need for the system to be observed at a given configuration and time. An intriguing question, then, concerns the origins of such a highly misaligned orbital architecture. One distinct possibility is a transient dynamical instability, that would have led to chaotic excitation orbital inclinations. Although such a scenario is not strictly impossible, the consistency of our RV fit with circular orbits renders such an evolutionary sequence unlikely. Some additional circumstantial evidence for long-term stability is the lack of a dense, hot disk around HD 3167, like that orbiting the G8V/K0V star HD 69830 \citep{Beichman2006}, which also hosts three planets \citep{Lovis2006}. Examining the WISE photometry \citep{Cutri2014} we find no evidence for an excess, which is expected for mature stars but the presence of which may be indicative of a recent disruptive event. 

An alternative, and perhaps more plausible solution is that the orbits have inherited their inclination from a primordially misaligned star. Over the past few years, theoretical evidence has been marshaled in support of the notion that stars can become misaligned with respect to their protoplanetary disks, during the T-Tauri stage of their lifetimes \citep{Bate2010,Lai2011,Batygin2012,Lai2014,Spalding2014,Matsakos2016}. An attractive feature of the primordial misalignment theory is that it can simultaneously account for the observed distribution of spin-orbit misalignments of hot Jupiters \citep{Spalding2015} as well as the inherent inclination dispersion \citep[often referred to as the \textit{Kepler} dichotomy][]{Ballard2016,Mazeh2015} of sub-Jovian planets \citep{Spalding2016}. Viewed in this context, HD 3167 probably represents an evolutionary outcome of a close-in planetary system that formed in a relatively quiescent environment, and was perturbed out of orbital alignment through secular exchange of angular momentum between the planets and the young star, while retaining orbital stability.

\section{Conclusions}

We have undertaken a large multi-site, multi-instrument campaign to characterize the masses of the planets in the bright, nearby system HD~3167. We find that the system is composed of a rocky super-Earth, a likely volatile-rich sub-Neptune, and discover a third, non-transiting planet. Using dynamical arguments we constrain the likely mutual inclination of the third planet to between 15--60 degrees, indicating a true mass which is also in the sub-Neptune range. Due to its high volatile component, HD~3167~c is a very promising target for \emph{HST} and \emph{JWST} characterization of its atmosphere. In particular, measuring the water content of the atmosphere could help inform whether the system, with its unique architecture, was formed {\it in situ}. Given the inherent difficulty in establishing comprehensive phase coverage for planets with orbital periods near to one day and one month, we emphasize the utility and necessity of collaborating across multiple RV instruments and sites in our analysis. HD~3167 is expected to be typical of the exoplanet systems discovered by the NASA \emph{TESS} mission: bright, late-type main-sequence host stars, likely hosting multiple small planets. As such, it illuminates some of the challenges involved in robust mass measurements of these systems, including the scope of the resources required to disentangle the system in the presence of additional non-transiting planets. This added expenditure of limited resources will need to be considered in the coordination and execution of the follow-up campaign for \emph{TESS} exoplanet targets. Given its location near the ecliptic plane, HD~3167 is in the maximum visibility window for the ESA CHEOPS mission \citep{Broeg2013}. This will allow for both the investigation of transit timing variations in planet c, and with improved knowledge of the orbit of planet d via ongoing radial velocity measurements, monitoring for potential transits of planet d.

\begin{deluxetable*}{lll}
\tablecaption{HD~3167 planet parameters.  t$_{14}$ is the total transit duration from the first to fourth contact. $S_{\rm{inc}}$ is the irradiation at the surface of the planet in units of the irradiation at Earth. For planet d, $T\rm{conj}_{d}$ is the time of inferior conjunction. \label{tab:pars}}
\tablehead{ 
    \colhead{Parameter}    & \colhead{Value}  & \colhead{Units}\\
}
\colnumbers
\startdata
Planet b &  & \\
\hline
Period & \periodb & days \\
Transit mid-point & \epochb  & BJD$_{\mathrm{TDB}}$ \\
$R_p/R_{\star}$ & \rponrsb & \\
$a/R_{\star}$ & \aonrsb & \\
$b$ & \impparb & \\
$i$ & \incb & deg \\
$e$ & 0 (fixed) & \\
Transit depth & \depthb & ppm\\
t$_{14}$ & \tdurb & hrs\\
$R_p$ & \pradb &  $R_{\oplus}$\\
$K$ & \rvampb & m s$^{-1}$\\
$M_p$ & \pmassberr & $M_{\oplus}$ \\
$\rho$ & \pdensberr & g cm$^{-3}$ \\
$a$ & \semiab & AU \\
$S_{\rm{inc}}$ & 1625$^{+244}_{-222}$ & $S_{\oplus}$ \\
\hline
Planet c & & \\
\hline
Period & \periodc & days \\
Transit mid-point & \epochc &  BJD$_{\mathrm{TDB}}$\\
$R_p/R_{\star}$ & \rponrsc & \\
$a/R_{\star}$ & \aonrsc & \\
$b$ & \impparc & \\
$i$ & \incc & deg \\
$e$ & $<$0.267 & \\
Transit depth & \depthc & ppm \\
t$_{14}$ & \tdurc & hr \\
$R_p$  &\pradc & $R_{\oplus}$\\
$K$ & \rvampc & m/s \\
$M_p$ & \pmasscerr & $M_{\oplus}$\\
$\rho$ & \pdenscerr & g cm$^{-3}$  \\
$a$ & \semiac & AU \\
$S_{\rm{inc}}$ & 16.6$^{+2.5}_{-2.3}$ & $S_{\oplus}$\\
\hline
Planet d &  & \\
\hline
Period & \periodd & days\\
$T\rm{conj}_{d}$ & \tcepochd & BJD$_{\mathrm{TDB}}$ \\
$e$ & $<$0.36 & \\
%$K$ & \rvampd & m s$^{-1}$\\
$M_p$ sin $i$ & \pmassderr & $M_{\oplus}$ \\
$a$ & \semiad & AU \\
$S_{\rm{inc}}$ & 88.9$\pm$6.2 & $S_{\oplus}$\\
\enddata
%\tablecomments{Note that {\tt \string \colnumbers} does not work with the 
%vertical line alignment token. If you want vertical lines in the headers you
%can not use this command at this time.}
\end{deluxetable*}

\acknowledgments

This paper and the paper by Gandolfi et al. were prepared simultaneously and are the result of independent radial-velocity observations and analyses of the HD~3167 system. We thank the HARPS team for their collegiality. We also thank the many observers who contributed to the measurements reported here. We thank Kyle Lanclos, Matt Radovan, Will Deich and the rest of the UCO Lick staff for their invaluable help shepherding, planning, and executing observations, in addition to writing the low-level software that made the automated APF observations possible. We are grateful to the time assignment committees of the University of Hawai'i, the University of California, and NASA for their generous allocations of observing time. A. W. H. acknowledges support for our K2 team through a NASA Astrophysics Data Analysis Program grant. A. W. H. and I. J. M. C. acknowledge support from the K2 Guest Observer Program. This material is based upon work supported by the National Science Foundation Graduate Research Fellowship under Grant No. 2014184874. Any opinion, findings, and conclusions or recommendations expressed in this material are those of the authors and do not necessarily reflect the views of the National Science Foundation. The research leading to these results has received funding from the European Union Seventh Framework Program (FP7/2007-2013) under grant agreement number 313014 (ETAEARTH). This publication was made possible through the support of a grant from the John Templeton Foundation. The opinions expressed are those of the authors and do not necessarily reflect the views of the John Templeton Foundation. This material is based upon work supported by NASA under grants No. NNX15AC90G and NNX17AB59G issued through the Exoplanets Research Program. Some of the data presented in this paper were obtained from the Mikulski Archive for Space Telescopes (MAST). STScI is operated by the Association of Universities for Research in Astronomy, Inc., under NASA contract NAS5-26555. Support for MAST for non-HST data is provided by the NASA Office of Space Science via grant NNX09AF08G and by other grants and contracts. This research has also made use of the NASA Exoplanet Archive, which is operated by the California Institute of Technology, under contract with the National Aeronautics and Space Administration under the Exoplanet Exploration Program. This research has made use of the NASA/IPAC Infrared Science Archive, which is operated by the Jet Propulsion Laboratory, California Institute of Technology, under contract with the National Aeronautics and Space Administration. The Digitized Sky Survey was produced at the Space Telescope Science Institute under U.S. Government grant NAG W-2166. The images of these surveys are based on photographic data obtained using the Oschin Schmidt Telescope on Palomar Mountain and the UK Schmidt Telescope. The plates were processed into the present compressed digital form with the permission of these institutions. This research has made use of the NASA Exoplanet Follow-Up Observation Program website, which is operated by the California Institute of Technology, under contract with the National Aeronautics and Space Administration under the Exoplanet Exploration Program. Finally, the authors wish to extend special thanks to those of Hawai'ian ancestry on whose sacred mountain of Maunakea we are privileged to be guests. Without their generous hospitality, the Keck observations presented herein would not have been possible.


%% for the paper.  Each keyword is check against the master list during
%% copy editing.  Individual instruments can be provided in parentheses,
%% after the keyword, but they are not verified.

\vspace{5mm}
\facilities{Kepler, Keck(HIRES, NIRC2), APF, HARPS-N}

\software{emcee \citep{Foreman-Mackey2013}, isochrones \citep{Morton2015}, RadVel (Fulton \& Petigura, in prep)}



%\appendix

%\section{Appendix information}

\listofchanges

\end{document}